\documentclass[12pt,onecolumn,twoside]{IEEEtran}
\usepackage[inline]{enumitem}
\usepackage{amsmath,mathrsfs,amssymb}
\usepackage{url}
\usepackage{setspace}
\usepackage[utf8]{inputenc}
\usepackage{bbm}
\usepackage{subcaption}
\usepackage{preamble}

\usepackage[english]{babel} %
\usepackage[autostyle=true]{csquotes}

\makeatletter
\let\l@ENGLISH\l@english
\makeatother

\usepackage[style=ieee, citestyle=numeric-comp, natbib=false, sorting=none,
  uniquename=false, hyperref=false, backend=bibtex, bibencoding=auto, 
  minnames=1,defernumbers=true, maxcitenames=2,
maxbibnames=99]{biblatex}

\AtEveryBibitem{\clearfield{doi}}
\AtEveryBibitem{\clearfield{issn}}
\AtEveryBibitem{\clearfield{isbn}}

\usepackage[activate={true,nocompatibility},final,tracking=true, 
kerning=true,spacing=true,factor=1100,stretch=10,shrink=10]{microtype}

\newtheorem{thm}{Theorem}

\newtheorem{cor}{Corollary}
\newtheorem{lem}{Lemma}
\newtheorem{rem}{Remark}

\newtheorem{assumption}{Assumption}

\usepackage{relsize}

\usepackage{xr}
\newcommand*{\putNumberT}[3]{
  \put(0,#2){
    \color{white}\small \makebox[0pt][r]{RSE=\SI{#1}{\percent}\rule{#3mm}{0pt}}
  }
}

\begin{document}

\title{Polychromatic X-ray CT Image Reconstruction and Mass-Attenuation 
Spectrum Estimation}
\author{\IEEEauthorblockA{Renliang Gu and Aleksandar Dogandžić\\
  ECpE Department, Iowa State University\\
  3119 Coover Hall, Ames, IA 50011\\ email:
  \texttt{\{renliang,ald\}@iastate.edu}}
  \thanks{This work was supported by the National Science Foundation under 
  Grant CCF-1421480.}
}

\maketitle
\vspace*{-2cm}
\begin{abstract}
  We develop a method for sparse image reconstruction from
  polychromatic \gls{CT} measurements under the blind scenario where the 
  material of the inspected object and the incident-energy spectrum are 
  unknown.  We obtain a parsimonious measurement-model parameterization by 
  changing the integral variable from photon energy to mass attenuation, 
  which allows us to combine the variations brought by the   unknown 
  incident spectrum and mass attenuation into a single unknown 
  \emph{mass-attenuation spectrum} function; the resulting measurement 
  equation has the Laplace integral form. The mass-attenuation spectrum is 
  then expanded into basis functions using B-splines of order one.  We 
  derive a block coordinate-descent algorithm for constrained minimization 
  of a penalized \gls{NLL} cost function, where penalty terms ensure 
  nonnegativity of the spline coefficients and nonnegativity and sparsity 
  of the density map image.  The image sparsity is imposed using convex 
  \gls{TV} and $\ell_1$ norms, applied to the density-map image and its 
  \gls{DWT} coefficients, respectively.  This algorithm alternates between 
  \gls{NPG} and \gls{LBFGSB} steps for updating the image and 
  mass-attenuation spectrum parameters.  To accelerate convergence of the 
  density-map \gls{NPG} step, we apply a step-size selection scheme that 
  accounts for varying local Lipschitz constant of the \gls{NLL}.  We 
  consider lognormal and Poisson noise models and establish conditions for 
  biconvexity of the corresponding \glspl{NLL} with respect to the density 
  map and  mass-attenuation spectrum parameters.  We also prove the 
  Kurdyka-{\L}ojasiewicz property of the objective function, which is 
  important for establishing local convergence of block-coordinate schemes 
  in biconvex optimization problems.  Numerical experiments with simulated 
  and real X-ray \gls{CT} data demonstrate the performance of the proposed 
  scheme.  \end{abstract}

\glsresetall

\section{Introduction}
\label{sec:intro}

X-ray \gls{CT} measurement systems are important in modern \gls{NDE} and 
medical diagnostics.  The past decades have seen great progress in \gls{CT} 
hardware and (reconstruction) software development.  \gls{CT} sees into the 
inspected object and gives 2D and 3D reconstruction at a high resolution.  
It is a fast high-resolution method that can distinguish a density 
difference as small as \SI{1}{\percent} between materials.  As it shows the 
finest detail of the inside, it has been one of the most important 
techniques in medical diagnosis, material analysis and characterization, 
and \gls{NDE} \cite{Dewulf2012,WangYuDeMan,Ying2006}.

Therefore, improving reconstruction accuracy and speed of data collection 
in these systems could have a significant impact on these broad areas.  
Thanks to recent computational and theoretical advances, such as 
\glspl{GPU} and sparse signal reconstruction theory and methods, it is now 
possible to design iterative reconstruction methods that incorporate 
accurate nonlinear physical models into sparse signal reconstructions from 
significantly undersampled measurements.

However, due to the polychromatic nature of the X-ray source, linear 
reconstructions such as \gls{FBP} exhibit beam-hardening artifacts, e.g., 
cupping and streaking \cite{Barrett2004}.
These artifacts limit the quantitative analysis of the reconstruction. In 
medical \gls{CT} application, severe artifacts can even look similar to 
certain pathologies and further mislead the diagnosis 
\cite[Sec.~7.6.2]{Hsieh2009CT}.  Fulfilling the promise of compressed 
sensing and sparse signal reconstruction in X-ray \gls{CT} depends on 
accounting for the polychromatic measurements as well. It is not clear how 
aliasing and beam-hardening artifacts interact and our experience is that 
we cannot achieve great undersampling when applying sparse linear 
reconstruction to polychromatic measurements.  Indeed, the error caused by 
the model mismatch may well be larger than the aliasing error that we wish 
to correct via sparse signal reconstruction.

\begin{figure*}
  \centering
  {\begin{subfigure}{0.55\textwidth}
    \centering
    \includegraphics{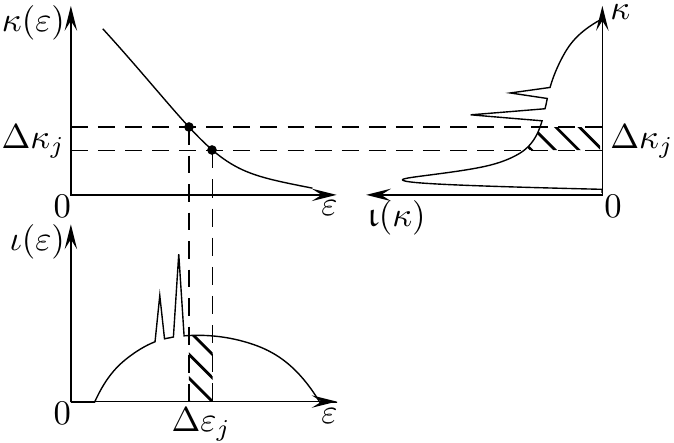}
    \caption{}
    \label{fig:triRelate}
  \end{subfigure}}
  {\begin{subfigure}{0.38\textwidth}
    \centering
    {\includegraphics[scale=0.90]{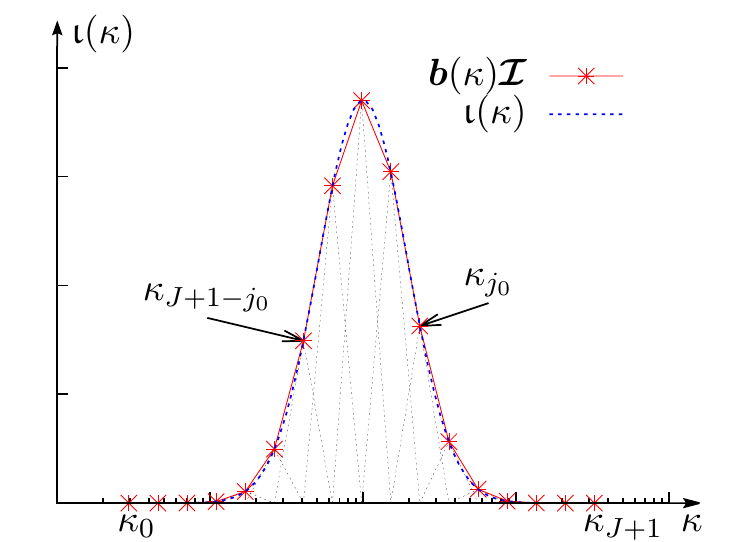}}
    \caption{}
    \label{fig:spline}
  \end{subfigure}}
  \caption{
    (a) Mass-attenuation spectrum $\upiota(\kappa)$ obtained by combining
    the mass attenuation $\kappa(\varepsilon)$ and incident
    spectrum $\iota(\varepsilon)$, and (b) its B1-spline expansion.
  }
\end{figure*}

In this paper (see also \cite{GuDogandzic2013,gdqnde14}), we adopt the 
nonlinear measurement scenario resulting from the polychromatic X-ray 
source and simplify the measurement model by exploiting the relation 
between the \emph{mass attenuation coefficients}, \emph{X-ray photon 
energy} and \emph{incident spectrum}, see Fig.~\ref{fig:triRelate}.  This 
simplified model allows \emph{blind} density-map reconstruction
and estimation of the composite \emph{mass attenuation spectrum} 
$\upiota(\kappa)$ (depicted in Fig.~\ref{fig:triRelate})
in the case where both the mass attenuation and incident spectrum are 
unknown.

We introduce the notation: $I_N$, $\bm{1}_{N \times 1}$, and $\bm{0}_{N
\times 1}$ are the  identity matrix of size $N$ and the $N \times 1$
vectors of ones and zeros (replaced by $I,\bm{1}$, and $\bm{0}$ when the 
dimensions can be inferred easily); $\bm{e}_n = I(:,n)$ is a column vector 
with the $n$th element equal to one and the remaining elements equal to 
zero; $\lvert \cdot \rvert$, $\lVert \cdot \rVert_p$, and ``$^T$'' are the 
absolute value, $\ell_p$  norm, and transpose, respectively.  Denote by
$\lceil x \rceil$  the smallest integer larger than
or equal to $x \in \mathbb{R}$.  For a vector $\balpha = [ \alpha_1, 
\dotsc, \alpha_p ]^T \in \mathbb{R}^p$, define
the  nonnegativity indicator function and projector
\begin{subequations}
\begin{IEEEeqnarray}{rCl}
  \label{eq:nonnegindf}
  \mathbb{I}_\nonneg(\balpha) &\df& \ccases{ 0, & \ba
    \succeq \bm{0}\\
    +\infty, & \text{otherwise}
  }\\
  \label{eq:nonnegproj}
  \SBR{(\ba)_+}_i &=& \maxp{a_i,0}
\end{IEEEeqnarray}
and the soft-thresholding operator 
$\SBR{\softthr{\lambda}{\balpha}}_i=\sign(\alpha_i)\maxp{ \abs{\alpha_i} 
-\lambda,0 }$, where ``$\succeq$'' is the elementwise version of 
``$\geq$''.  Furthermore, $\ba^\tL(s) \df \int \ba(\kappa)\E^{-s\kappa}\dif 
\kappa$ is the \emph{Laplace transform} of a vector function $\ba(\kappa)$ 
and
 \begin{IEEEeqnarray}{c"c}
  \label{eq:LTders}
  \PARENS{ (-\kappa)^m \ba}^\tL(s) = \int (-\kappa)^m 
  \ba(\kappa)\E^{-s\kappa}\dif \kappa = \od[m]{\ba(s)}{s}, & m \in 
  \mathbb{N}_0.
\end{IEEEeqnarray}
\end{subequations}
Define also the set of  nonnegative real numbers as $\mathbb{R}_+=\nonneg$, 
the elementwise logarithm $\ln_{\circ}(\bx) = [ \ln x_1,  \ldots, \ln x_N 
]^T$, nonnegativity projector $[(\bx)_+]_i=\max\{x_i,0\}$  where $\bx = 
[x_1, x_2, \dotsc, x_N ]^T$, and Laplace transforms $\ba_\circ^\tL(\bs) =  
\PARENS{ \ba^\tL(s_n) }_{n=1}^N$ and $(\kappa\ba)_\circ^\tL(\bs) = \PARENS{ 
(\kappa\ba)^\tL(s_n) }_{n=1}^N$ obtained by stacking $\ba^\tL(s_n)$ and 
$(\kappa\ba)^\tL(s_n)$ columnwise,  where $\bs = [ s_1, s_2, \dotsc, s_N 
]^T$.  Finally, $\supp(\iota(\cdot))$ is the support set of a function 
$\iota(\cdot)$, $\dom(f)=\{\bx\in\mathbb{R}^n\given{f(\bx)<+\infty}\}$ is 
the domain of function $f(\cdot)$, and $\diagp{\bx}$ is the diagonal matrix 
with diagonal elements defined by the corresponding elements of vector 
$\bx$.

In the following, we review the standard noiseless polychromatic X-ray CT 
measurement model (Section~\ref{sec:polyCTmodel}), existing signal 
reconstruction approaches from polychromatic measurements 
(Section~\ref{sec:existingpolyCT}), and our mass-attenuation 
parameterization of this model (Section~\ref{sec:massatennuationparam}).

\subsection{Polychromatic X-ray CT Model}
\label{sec:polyCTmodel}

By the \emph{exponential law of absorption} \cite{JenkinsWhite}, the
fraction $\dif \cI / \cI$ of plane wave intensity lost in traversing
an infinitesimal thickness $\dif\ell$ at Cartesian coordinates $(x,y)$
is proportional to $\dif\ell$:
\begin{equation}
  \frac{\dif \cI}{\cI} = - \mu(x,y,\varepsilon) \dif\ell
  \label{eq:beerLaw1}
\end{equation}
where $\mu(x,y,\varepsilon)$ is the attenuation.  To obtain the
intensity decrease along a straight-line path $\ell=\ell(x,y)$ at
photon energy $\varepsilon$, integrate \eqref{eq:beerLaw1} along
$\ell$: $\cI^{\text{out}}(\varepsilon) = \cI^{\text{in}}(\varepsilon)
\exp \SBR{-{\int_\ell \mu(x,y,\varepsilon)\dif\ell}}$, where
$\cI^{\text{out}}(\varepsilon)$ and $\cI^{\text{in}}(\varepsilon)$ are
the emergent and incident X-ray signal energies, respectively, at
photon energy $\varepsilon$.

To describe the polychromatic X-ray source, assume that its incident
intensity $\cI^{\text{in}}$ spreads along photon energy $\varepsilon$
following the density $\iota(\varepsilon) \geq 0$, i.e.,
\begin{subequations}
  \label{eq:continuousModel}
  \begin{equation}
    \label{eq:constanttotalenergy}
    \int \iota(\varepsilon) \dif\varepsilon = \cI^{\text{in}}.
  \end{equation}
  see Fig.~\ref{fig:triRelate}, which shows a typical density
  $\iota(\varepsilon)$.
  The noiseless measurement collected by an energy-integrating detector
  upon traversing a straight line $\ell=\ell(x,y)$ has the 
  superposition-integral form:
  \begin{IEEEeqnarray}{rCl}
    \cI^{\text{out}} &=& \int \iota(\varepsilon) \exp\SBR{- {\int_\ell
    \mu(x,y,\varepsilon) \dif\ell}} \dif\varepsilon \notag
    \\
    &=& \int \iota(\varepsilon) \exp\SBR{- \kappa(\varepsilon)
    \int_\ell \alpha(x,y) \dif\ell} \dif\varepsilon
    \label{eq:polyModelseparable}
  \end{IEEEeqnarray}
\end{subequations}
where we model the attenuation $\mu(x,y,\varepsilon)$ of the inspected 
object consisting of a single material using the following \emph{separable 
form} \cite[Sec.~6]{Nuyts2013}:
\begin{equation}
  \label{eq:museparable}
  \mu(x,y,\varepsilon) = \kappa(\varepsilon) \alpha(x,y).
\end{equation}
Here, $\kappa(\varepsilon)>0$ is the \emph{mass-attenuation coefficient} of 
the material, a function of the photon energy $\varepsilon$ (illustrated in 
Fig.~\ref{fig:triRelate}) and $\alpha(x,y) \geq 0$ is the density map of 
the object. For a monochromatic source at photon energy $\varepsilon$, 
$\ln[\cI^{\text{in}}(\varepsilon)/\cI^{\text{out}}(\varepsilon)]$ is a 
linear function of $\alpha(x,y)$, which is a basis for traditional linear 
reconstruction.  However, X-rays generated by vacuum tubes are not 
monochromatic \cite{Kak1988CT,Hsieh2009CT} and we cannot transform the 
underlying noiseless measurements to a linear model \emph{unless} we know 
perfectly the incident energy spectrum $\iota(\varepsilon)$ and mass 
attenuation of the inspected material $\kappa(\varepsilon)$.\looseness=-1

\subsection{Existing approaches for signal reconstruction from 
polychromatic X-ray CT measurements}
\label{sec:existingpolyCT}

Beam-hardening correction methods can be categorized into
pre-filtering, linearization, dual-energy, and post-reconstruction
approaches \cite{Krumm2008}.
Reconstruction methods have recently been
developed in \cite{Fessler2002StatisticalBH, ElbakriFessler2003,
  Nuyts2011} that aim to optimize nonlinear objective functions based
on the underlying physical model;
\cite{Fessler2002StatisticalBH,ElbakriFessler2003} assume known
incident polychromatic source spectrum and imaged materials, whereas
\cite{Nuyts2011} considers a blind scenario with \emph{unknown}
incident spectrum and imaged materials, but employs an excessive
number of parameters and suffers from numerical instability. See also
the discussion below and in \cite{gdCAMSAP13}.

\subsubsection{Linearization}  \label{sec:linearization}
Observe that \eqref{eq:polyModelseparable} is
a \emph{decreasing function} of the line integral $\int_\ell
\alpha(x,y) \dif\ell$, which is a key insight behind linearization methods 
and the water correction approach in
\cite{Herman1979}, popular in the literature and practice, see
\cite[Sec.~6]{Nuyts2013}.  The linearization function has been
approximated by exponential or polynomial functions
\cite{Huang2009,Kingston2012}.  The linearization
functions are obtained by a variety of approaches:
\cite{Herman1979,Huang2009} assume known spectrum and the materials to
simulate the relation between the polychromatic and monochromatic
projections; \cite{Kingston2012} aims to maximize the
self-consistency of the projection data to get the correction curve
blindly; \cite{Krumm2008} segments the initial reconstruction to
obtain propagation path length of X-ray and then determines the
linearization curve.

Indeed, we can derive analytically [if $\kappa(\varepsilon)$ and
$\iota(\varepsilon)$ are known] or estimate this monotonic nonlinear
function and apply the corresponding inverse function to the noisy
measurements. However, such a zero-forcing approach to reconstruction
ignores noise and therefore leads to noise enhancement.  Due to the
difficulty in getting precise linearization function, it has also been
reported to introduce a dependency of dimensional measurements on the
amount of surrounding material, which is a serious drawback in
applications such as CT metrology \cite{Dewulf2012}. The scanning
configuration, e.g., X-ray power and prefiltrations, makes choosing
accurate linearization function harder.  The mass-attenuation
parameterization in Fig.~\ref{fig:triRelate}
 will allow for improved selection
and estimation of the linearization function compared with the
existing work. Unlike the traditional linearization, the methods that
we aim to develop will estimate the linearization function \emph{and}
account for the noise effects. The zero-forcing linearization schemes
could be used to initialize our iterative procedures.

\subsubsection{Post-reconstruction}
The post-reconstruction approach first proposed in \cite{Nalcioglu1979}, 
uses the segmentation of an initial reconstruction to estimate the 
traversal length of each X-ray through each material. It assumes known 
incident X-ray spectrum and attenuation coefficients, as well as that the 
imaged materials are homogeneous and have constant density, leading to 
piecewise-constant reconstructed images.

\subsubsection{Poisson and lognormal statistical measurement models}
X-ray photons arriving at a detector can be well modeled using a Poisson 
process \cite{Nuyts2001BH,ElbakriFessler2003}, where the energy deposited 
by each photon follows a distribution that is proportional to the incident 
energy spectrum.  Hence, the measurements collected by photon-counting and 
energy-integrating detectors are modeled using Poisson and compound-Poisson 
distributions, respectively 
\cite{Elbakri2003Compound,Whiting2006Properties}.  However, the 
compound-Poisson distribution is complex and does not have a closed-form 
expression, which motivates introduction of approximations 
such as Poisson \cite{Fessler2002StatisticalBH} and lognormal 
\cite{Nuyts2011,Xu2014,gdqnde14}.  Regardless of the statistical 
measurement models, most most methods in this category
assume known X-ray spectrum and materials (i.e., known mass-attenuation 
function), with goal to maximize the underlying likelihood function or its 
regularized version \cite{Fessler2002StatisticalBH, Evans2013Experimental}.  
However, the X-ray spectrum measurements based on the semiconductor 
detectors are usually distorted by charge trapping, escape events, and 
other effects \cite{Redus2009} and the corresponding correction requires 
highly collimated beam and special procedures \cite{Zhang2012, Lin2014}.  
Knowing the mass-attenuation function can be challenging as well when the 
inspected material is unknown, or the inspected object is made of compound 
or mixture with unknown percentage of each constituent.

\citeauthor{Nuyts2011} \cite{Nuyts2011} consider a ``blind'' scenario
for lognormal measurement model
with \emph{unknown} incident spectrum and materials %
and employ the $K$-means clustering method to initially associate pixels to 
the materials and then alternate between material segmentation and updating 
the relative density map, incident X-ray spectrum, and mass attenuation 
coefficients for each material.  The methods in \cite{Nuyts2011} are based 
on the standard photon-energy parameterization in 
\eqref{eq:continuousModel}, employ an excessive number of parameters, and 
suffer from numerical instability \cite{gdCAMSAP13}.  Indeed, alternating 
iteratively between updating excessive numbers of non-identifiable 
parameters prevents development of elegant algorithms and achieving robust 
reconstructions.

\subsubsection{Dual-energy} In the dual-energy approach by 
\citeauthor{Alvarez1976co} \cite{Alvarez1976co}, two X-ray scans are 
performed using two different incident X-ray spectra. This approach is 
particularly attractive because it does not require image segmentation to 
separate out different materials and it provides two reconstruction images.  
Depending on the basis functions chosen for modeling the attenuation 
coefficients, the two reconstructed images depict two coefficient maps 
representing photoelectric and Compton scattering interaction 
\cite{Alvarez1976co} or two density maps of the chosen basis materials 
\cite{ZhangBouman2014}.  The dual-energy approach doubles the scanning 
time, needs switching scanning energy, and requires the exact knowledge of 
the spectra.

\subsection{Mass-attenuation parameterization}
\label{sec:massatennuationparam}

We now introduce our parsimonious parameterization of
\eqref{eq:polyModelseparable} for signal reconstruction.  Since the
mass attenuation $\kappa(\varepsilon)$ and incident spectrum density
$\iota(\varepsilon)$ are both functions of $\varepsilon$ (see
Fig.~\ref{fig:triRelate}), we combine the variations of these two
functions and write \eqref{eq:constanttotalenergy} and
\eqref{eq:polyModelseparable} as integrals of $\kappa$ rather than
$\varepsilon$, with goal to represent our model using two functions
$\upiota(\kappa)$ (defined below) and $\alpha(x,y)$ instead of three
[$\iota(\varepsilon), \kappa(\varepsilon)$, and $\alpha(x,y)$], see
also \cite{GuDogandzic2013}.
Hence, we rewrite \eqref{eq:constanttotalenergy} and 
\eqref{eq:polyModelseparable} as (see Appendix~\ref{app:anykappa})
\begin{subequations}
  \label{eq:massattenuationpar}
  \begin{IEEEeqnarray}{rCl}
    \cI^{\textup{in}} &=& \upiota^\tL(0)\\
    \cI^{\mathrm{out}} &=& \upiota^\tL\PARENS{\int_\ell\alpha(x,y)\dif\ell}
  \end{IEEEeqnarray}
\end{subequations}
where $\upiota^\tL(s) = \int \upiota(\kappa)\E^{-s\kappa}\dif \kappa$ is 
the Laplace transform of
the \emph{mass-attenuation spectrum}  $\upiota(\kappa)$, which represents 
the density of the incident X-ray energy at attenuation $\kappa$; here, 
$s>0$, in contrast with the traditional Laplace transform where $s$ is 
generally complex.  For invertible $\kappa(\varepsilon)$ with 
differentiable inverse
function $\varepsilon(\kappa)$, $\upiota(\kappa) =
\iota(\varepsilon(\kappa)) |\varepsilon'(\kappa)| \geq 0$, with
$\varepsilon'(\kappa) = \dif \varepsilon(\kappa) / \dif \kappa$, see 
Appendix~\ref{app:invertiblekappa}.  All
$\kappa(\varepsilon)$ encountered in practice can be divided into
$M+1$ piecewise-continuous segments $\{(e_m,e_{m+1})\}_{m=0}^M$, where 
$\kappa(\varepsilon)$ in  each segment is a differentiable
monotonically decreasing function of $\varepsilon$. In this case,
\eqref{eq:massattenuationpar} holds and the expression for
$\upiota(\kappa)$ can be easily extended to this scenario, yielding
(see Appendix~\ref{app:piecewisecontinuouskappa})
\begin{equation}
  \upiota(\kappa) = \sum_{m=0}^M 1_{(u_m, v_m)}(\kappa)
  \iota(\varepsilon_m(\kappa)) \left| \varepsilon_m'(\kappa) \right|
  \label{eq:generalUpiota}
\end{equation}
where $1_{(u_m, v_m)}(\kappa)$ is an indicator function that takes value 1 
when $\kappa\in(u_m, v_m)$ and 0 otherwise.
The range and inverse of $\kappa(\varepsilon)$ within $(e_m, e_{m+1})$ are 
$(u_m, v_m)$ and $\varepsilon_m(\kappa)$, respectively, with
$u_m\df\inf_{\varepsilon{\scriptscriptstyle\nearrow}e_{m+1}}\kappa(\varepsilon)
<
v_m\df\sup_{\varepsilon{\scriptscriptstyle\searrow}e_m}\kappa(\varepsilon)$.

Observe that the mass-attenuation spectrum $\upiota(\kappa)$ is nonnegative 
for all $\kappa$:
\begin{IEEEeqnarray}{c"c}
  \label{eq:analogconstraints}
  \upiota(\kappa) \geq 0.
\end{IEEEeqnarray}
Due to its nonnegative support and range, $\upiota^\tL(s)$ is a decreasing 
function of $s$.  Here, $s>0$, in contrast with the traditional Laplace 
transform where $s$ is generally complex.  The function 
$(\upiota^\tL)^{-1}$ converts the noiseless measurement $\cI^{\tout}$ in 
\eqref{eq:massattenuationpar}, which is a nonlinear function of the density 
map $\alpha(x,y)$,  into a noiseless linear ``measurement'' $\int_\ell 
\alpha(x,y) \dif \ell$.  The $(\upiota^\tL)^{-1}\circ\exp(-\cdot)$ mapping 
corresponds to the \emph{linearization function} in \parencite{Herman1979} 
[where it was defined through \eqref{eq:polyModelseparable}  rather than 
the mass-attenuation spectrum] and converts $-{\ln \cI^{\tout}}$ into a  
noiseless linear  ``measurement'' $\int_\ell \alpha(x,y) \dif \ell$, see 
also the discussion in Section~\ref{sec:linearization}.

The mass-attenuation spectrum depends both on the measurement system 
(through the incident energy spectrum) and inspected object (through the 
mass attenuation of the inspected material).  In the blind scenario where 
the inspected material and incident signal spectrum are unknown, the above 
parameterization allows us to estimate two functions [$\upiota(\kappa)$ and 
$\alpha(x,y)$] rather than three [$\iota(\varepsilon), 
\kappa(\varepsilon)$, and $\alpha(x,y)$].  This blind scenario is the focus 
of this paper.

In Section~\ref{sec:meaModel}, we introduce the density-map and 
mass-attenuation parameters to be estimated
and discuss their identifiability.
Section~\ref{sec:measurementmodel} then presents lognormal and Poisson 
probabilistic measurement models and biconvexity properties of their 
\glspl{NLL}.   Section~\ref{sec:estimation} introduces the penalized 
\gls{NLL} function that incorporates the parameter constraints and 
describes the proposed block coordinate-descent algorithm, which alternates 
between a \gls{NPG} step for estimating the density map image and a 
\gls{LBFGSB} step for estimating the incident-spectrum parameters.  In 
Section~\ref{sec:numex}, we show the performance of the proposed method 
using simulated and real X-ray \gls{CT} data. Concluding remarks are given 
in Section~\ref{sec:conclusion}.

\section{Basis-Function Expansion of Mass-Attenuation Spectrum}
\label{sec:meaModel}

Upon spatial-domain discretization into $p$ pixels,  replace the integral 
$\int_\ell \alpha(x,y) \dif \ell$ with $\bphi^T \balpha$:
\begin{IEEEeqnarray}{c}
  \label{eq:discroverspace}
  \int_\ell \alpha(x,y) \dif \ell \approx \bphi^T \balpha
\end{IEEEeqnarray}
where $\balpha \succeq \bm{0}$ is a $p \times 1$  vector representing the 
2D image that we wish to reconstruct [i.e., discretized $\alpha(x,y)$, see 
also \eqref{eq:analogconstraints}] and $\bphi \succeq \bm{0}$ is a $p 
\times 1$ vector of known weights quantifying how much each element of 
$\balpha$ contributes to the X-ray attenuation on the straight-line path 
$\ell$.  An X-ray \gls{CT} scan consists of hundreds of projections with the 
beam intensity measured by thousands of detectors for each projection.  Denote 
by $N$ the total number of measurements from all projections collected at the 
detector array.
For the $n$th measurement, define its discretized line integral as $\bphi_n^T 
\balpha$; stacking all $N$ such integrals into a vector yields $\Phi\balpha$, 
where $\Phi= \SBR{\bphi_1 \, \bphi_2 \dotsb \bphi_N }^T\in\mathbb{R}^{N\times 
p}$ is the \emph{projection matrix},   also called Radon transform matrix in a 
parallel-beam X-ray tomographic imaging system.  We call the corresponding 
transformation, $\Phi\balpha$, the \emph{monochromatic projection} of 
$\balpha$.\looseness=-1

Approximate $\upiota(\kappa)$ with a linear combination of $J$ ($J \ll N$) 
basis functions:
\begin{subequations}
\begin{IEEEeqnarray}{c}
  \label{eq:upiota}
  \upiota(\kappa) = \bb(\kappa) \bcI
\end{IEEEeqnarray}
where \begin{IEEEeqnarray}{rCl}
  \label{eq:I}
\bcI&\df&[ \mathcal{I}_1, \mathcal{I}_2, \dotsc, \mathcal{I}_J]^T \\
&\succeq& \bm{0}
\label{eq:Inonneg}
\end{IEEEeqnarray}
is the $J \times 1$ vector of corresponding basis-function coefficients and  
the $1 \times J$ row vector function \begin{IEEEeqnarray}{c}
  \label{eq:bb}
  \bb(\kappa) \df \SBR{ b_1(\kappa), b_2(\kappa),\dotsc, b_J(\kappa)}
\end{IEEEeqnarray}
\end{subequations}
consists of B-splines \cite{Schumaker2007Spline}  of order one  (termed B1 
splines, illustrated in Fig.~\ref{fig:spline}); in this case, the 
decomposition \eqref{eq:upiota} yields nonnegative elements of the spline 
coefficients $\bcI$ [based on \eqref{eq:analogconstraints}] and thus allows 
us to impose the physically meaningful nonnegativity constraint 
\eqref{eq:Inonneg} when estimating $\bcI$.  The spline knots are selected 
from a growing geometric series $\{\kappa_j\}_{j=0}^{J+1}$ with $\kappa_0 > 
0$,
\begin{subequations}
\begin{IEEEeqnarray}{c}
    \label{eq:kappaj}
  \kappa_j=q^j\kappa_0
\end{IEEEeqnarray}
and common ratio
\begin{IEEEeqnarray}{c}
  \label{eq:q}
  q>1
\end{IEEEeqnarray}
which yields the B1-spline basis functions:
\begin{IEEEeqnarray}{c}
  b_j(\kappa) =\ccases{
    \displaystyle   \frac{\kappa-\kappa_{j-1}}{\kappa_j-\kappa_{j-1}},
    & \kappa_{j-1}\leq\kappa<\kappa_j \\
    \displaystyle
    \frac{-\kappa+\kappa_{j+1}}{\kappa_{j+1}-\kappa_j},
    & \kappa_j\leq\kappa<\kappa_{j+1} \\
    0, & \text{otherwise}
  }
  \label{eq:bj}
\end{IEEEeqnarray}
where the $j$th basis function can be obtained simply by
$q$-scaling the $(j-1)$th basis function:
\begin{equation}
  \label{eq:bfscalingjjminus1}
  b_j(\kappa) =   b_{j+1}\PARENS{q\kappa}
\end{equation}
see also Fig.~\ref{fig:spline}.

The geometric-series knots have a wide span from $\kappa_0$ to 
$\kappa_{J+1}$ and compensate larger $\kappa$ (i.e., higher attenuation 
implying ``exponentially'' smaller $\E^{-\bphi^T\balpha\kappa_j}$ 
term\footnote{
  While using $\bb^\tL(\cdot)\bcI$ to approximate 
  $\upiota^\tL(\cdot)=\int\upiota(\kappa)\E^{-\cdot\kappa}\dif\kappa$, to 
  balance the weight to each $\{\cI_j\}_{j=1}^J$, we prefer 
  $\{b_j^\tL(\cdot)\}_{j=1}^J$ that can provide similar values over $j$.
  This desired property is achieved through the geometric-series knots.  }) 
  with a ``geometrically'' wider integral range, that results in a more 
  effective approximation of \eqref{eq:massattenuationpar}.  The common 
  ratio $q$ in \eqref{eq:q} determines the resolution of the B1-spline 
  approximation.  Here, we select $q$ and $J$ so that the range of $\kappa$ 
  spanning the mass-attenuation spectrum is constant:
\begin{IEEEeqnarray}{c}
  \label{eq:qJcond}
  \frac{\kappa_{J+1}}{\kappa_0} = q^{J+1}=\text{const}. 
\end{IEEEeqnarray}
In summary, the following three tuning constants:
\begin{IEEEeqnarray}{c}
  \label{eq:B1tuning}
  (q, \kappa_0, J)
\end{IEEEeqnarray}
\end{subequations}
define our B1-spline basis functions $\bb(\kappa)$.

Substituting \eqref{eq:discroverspace} and \eqref{eq:upiota} into 
\eqref{eq:massattenuationpar} for each of the $N$ measurements yields the 
following expressions for the incident energy and the $N \times 1$ vector of 
noiseless measurements:
\begin{subequations}
\begin{IEEEeqnarray}{rCl}
  \cI^{\textup{in}}(\bcI) &=& \bb^\tL(0) \bcI \\
  \bcI^\tout(\balpha,\bcI) &=& \bb_\circ^\tL(\Phi\balpha)\bcI
  \label{eq:IinIoutparI}
\end{IEEEeqnarray}
\end{subequations}
where, following the notation in Section~\ref{sec:intro}, $\b{b}_\circ^\tL( 
\bs) = \PARENSbig{ \bb^\tL(s_n ) }_{n=1}^N$ is an \emph{output 
basis-function matrix} obtained by stacking the $1 \times J$ vectors 
$\bb^\tL( s_n )$ columnwise, and $\bs = \Phi \balpha$ is the monochromatic 
projection.  Since the Laplace transform of \eqref{eq:bj} can be computed 
analytically, $\bb^\tL( s )$ has a closed-form expression.

\subsection{Ambiguity of the density map and mass-attenuation spectrum}
\label{sec:ambiguity}

We discuss density-map scaling ambiguity under the blind scenario where 
both the density map $\balpha$ and incident spectrum parameters $\bcI$ are 
unknown.  By noting \eqref{eq:bfscalingjjminus1} and the $\kappa$-scaling 
property of the Laplace transform,
$b_j\PARENS{q{\kappa}} \stackrel{\mathcal{L}}{\rightarrow} 
\frac{1}{q}b_{j}^\tL(s/q)$ for $q>0$, we conclude that selecting $q$ times 
narrower basis functions
$\SBR{ b_0(\kappa), b_1(\kappa), \dotsc, b_{J-1}(\kappa) }$ than those in
$\bb(\kappa)$ and $q$ times larger density map and spectral
parameters ($q \balpha$ and $q \bcI$) yields the same
mean output photon energy. Consequently,
\begin{IEEEeqnarray}{rCl}
  \bcI^\tout\bigl(\balpha, [0,\cI_2,\dotsc,\cI_J]^T \bigr) &=& 
  \bcI^\tout\bigl(q  \balpha, q [\cI_2,\dotsc,\cI_J,0]^T \bigr).
  \label{eq:shiftambiguity}
\end{IEEEeqnarray}
Hence, when $\bcI$ has a leading zero, the noiseless signal output 
\emph{remains the same} if we
shift the elements of $\bcI$ (and correspondingly the mass-attenuation 
spectrum) to the left, followed by scaling the new $\bcI$ and $\balpha$ by 
$q$. Equivalently, when $\bcI$ has a trailing zero, the noiseless signal 
output remains the same if we
shift the elements of $\bcI$ (and correspondingly the spectrum) to the 
right and scale the new $\bcI$ and $\balpha$ by $1/q$.  We refer to this 
property as the \emph{shift ambiguity} of the mass-attenuation spectrum, which 
allows us to rearrange leading or trailing zeros in the mass-attenuation 
coefficient vector $\bcI$ and position the central nonzero part of $\bcI$.

In Section \ref{sec:measurementmodel},
we introduce the lognormal and Poisson measurement models 
(Sections~\ref{sec:lognormal} and \ref{sec:Poisson}) and establish 
conditions for biconvexity of the corresponding \glspl{NLL} with respect to 
the density map and incident-spectrum parameters 
(Section~\ref{sec:biconvexity}).

\subsection{Rank of $\bb_\circ^\tL(\Phi\balpha)$ and selection of the 
number of splines $J$}
\label{sec:rankofA}

If $\bb_\circ^\tL(\Phi\balpha)$ does not have full column rank, then $\bcI$ 
is not identifiable even if $\balpha$ is known, see \eqref{eq:IinIoutparI}.  
The estimation of $\bcI$ may be numerically unstable if  
$\bb_\circ^\tL(\Phi\balpha)$ is poorly conditioned and has small minimum 
singular values.

We can think of the noiseless X-ray CT measurements as $\bb^\tL(s)\bcI$
sampled at different $s =  \bphi^T_n\balpha \in 
\SBRbig{0,\max_n(\bphi^T_n\balpha)}$. The following lemma implies that, if 
we could collect all $s\in[0,a],\,a>0$ (denoted $\bs$), then the 
corresponding $\bb_\circ^\tL(\bs)$ would be a full-rank matrix.

\begin{lem}
  \label{th:independence}
  $\bcJ=\bm{0}_{J\times1}$ is the necessary condition for 
  $\bb^\tL(s)\bcJ=0$ over the range $s\in[0,a]$, where 
  $\bcJ\in\mathbb{R}^J$ and $a>0$.
\end{lem}
\begin{IEEEproof}
  Use the fact that $\bb^\tL(s)\bcJ$ is analytic with respect to $s$.  If 
  $\bb^\tL(s)\bcJ=0$ in the domain $s \in [0,a]$, the $n$th order 
  derivative of $\bb^\tL(s)\bcJ$ with respect to $s$ is zero over this 
  domain for all $n=1,2,\ldots$.  Then, according to the Taylor expansion 
  of the whole complex plane at some point in $[0,a]$, 
  $\bb^\tL(\cdot)\bcJ=0$ everywhere.  Therefore, $\bb(\kappa)\bcJ=0$, thus, 
  $\bcJ=\bm{0}$.
\end{IEEEproof}

If our data collection system can sample over 
$\SBRbig{0,\max_n(\bphi^T_n\balpha)}$ sufficiently densely, we expect
$\bb_\circ^\tL(\Phi\balpha)$ to have full column rank.

As the number of splines $J$ increases for fixed support 
$[\kappa_0,\kappa_{J+1}]$ [see \eqref{eq:qJcond}], we achieve better 
resolution of the mass-attenuation spectrum, but 
$\bb_\circ^\tL(\Phi\balpha)$ becomes poorly conditioned with
its smallest singular values approaching zero.  To estimate this spectrum 
well, we wish choose a $J$ that provides both good resolution \emph{and} 
sufficiently large smallest singular value of $\bb_\circ^\tL(\Phi\balpha)$.  

Fortunately, we focus on the reconstruction of $\balpha$, which is affected 
by $\bcI$ only through the function $\bb^\tL(s)\bcI$, and $\bb^\tL(s)\bcI$ 
is stable as we increase $J$. Indeed, we observe that, when we choose $J$  
that is significantly larger than the rank of $\bb_\circ^\tL(\Phi\balpha)$, 
the estimation of $\balpha$ will be good and $\bb_\circ^\tL(\Phi\balpha)$ 
stable, even though the estimation of $\bcI$ is poor due to its 
non-identifiability. The increase of $J$ will also increase the 
computational complexity of signal reconstruction under the blind scenario 
where the mass-attenuation spectrum is unknown.

\section{Measurement Models and Their Properties}
\label{sec:measurementmodel}

\subsection{Lognormal measurement model}
\label{sec:lognormal}

Consider an $N \times 1$ vector $\bcE$ of energy measurements corrupted by 
independent lognormal noise and the corresponding \gls{NLL} function [see 
\eqref{eq:IinIoutparI}]:
\begin{IEEEeqnarray}{rCl}
  \label{eq:lognormalNLL}
  \cL\PARENS{\balpha,\bcI} = \frac{1}{2}\bigl\|
    \ln_\circ\bcE -\ln_\circ \bcI^\tout(\balpha,\bcI) \bigr\|_2^2.
\end{IEEEeqnarray}

In the following, we express the \gls{NLL} \eqref{eq:lognormalNLL} as a 
function of $\balpha$ with $\bcI$ fixed (Section~\ref{sec:NLLalpha}) and 
vice versa (Section~\ref{sec:NLLI}), and derive conditions for its 
convexity under the two scenarios.  These results will be used to establish 
the biconvexity conditions for this \gls{NLL} in 
Section~\ref{sec:biconvexity} and to describe our block coordinate-descent 
estimation algorithm in Section~\ref{sec:estimation}.

\subsubsection{NLL of $\balpha$}
\label{sec:NLLalpha}
\begin{subequations}
  Recall \eqref{eq:upiota} and define
  \begin{equation}
    \upiota_\circ^\tL(\Phi\balpha) = \bb_\circ^\tL(\Phi \balpha)\bcI
  \end{equation}
  obtained by stacking $\CBRbig{\upiota^\tL(\bphi_n^T\balpha)}_{n=1}^N$ 
  columnwise.
  The \gls{NLL} of $\balpha$ for fixed $\bcI$ is
  \begin{IEEEeqnarray}{rCl}
    \label{eq:lognormalNLLalpha}
    \cL_\upiota(\balpha) = \frac{1}{2} \bigl\|
      \ln_\circ \bcE  - \ln_\circ \upiota_\circ^\tL(\Phi \balpha) 
      \bigr\|_2^2
  \end{IEEEeqnarray}
  which corresponds to the Gaussian \gls{GLM} for measurements 
  $\ln_\circ(\bcE)$ with design matrix $\Phi$ and link function 
  $(\upiota_\circ^\tL)^{-1}\PARENS{\exp(\cdot)}$.
See \parencite{McCullagh1989} for introduction to \glspl{GLM}.
\end{subequations}

To establish convexity of the \gls{NLL} \eqref{eq:lognormalNLLalpha}, we 
enforce monotonicity of the mass-attenuation spectrum $\upiota(\kappa)$ in 
low- and high-$\kappa$ regions and also assume that the mid-$\kappa$ region 
has higher spectrum than the low-$\kappa$ region.
Note that we do not require here that $\upiota(\kappa)$ satisfies the 
basis-function expansion \eqref{eq:upiota}; however, the basis-function 
expansion requirement will be needed to establish the biconvexity of the 
\gls{NLL} in \eqref{eq:lognormalNLL}.
Hence, we define the three $\kappa$ regions using the spline parameters 
\eqref{eq:B1tuning} as well as an additional integer constant 
\begin{subequations}
\begin{IEEEeqnarray}{c}
  \label{eq:j0}
  j_0\geq\lceil(J+1)/2\rceil.
\end{IEEEeqnarray}
In particular, the low-, mid-, and high-$\kappa$ regions are defined as 
\begin{IEEEeqnarray}{rCl"rCl"rCl}
  \label{eq:Kapparegions}
  \mathcal{K}_\text{low} &\df& [\kappa_0,\kappa_{J+1-j_0}], &
  \mathcal{K}_\text{mid} &\df& [\kappa_{J+1-j_0},\kappa_{j_0}] &
  \mathcal{K}_\text{high} &\df& [\kappa_{j_0},\kappa_{J+1}].
\end{IEEEeqnarray}
\end{subequations}

\begin{assumption}
  \label{assumption:I}
  Assume that the mass-attenuation spectrum satisfies
  \begin{IEEEeqnarray}{rl}
    \label{eq:upiotaShape}
    \mathcal{A}=\biggl\{
      \lgiven{
        \upiota:\;[\kappa_0,\kappa_{J+1}]\rightarrow\mathbb{R}_+
      }&
      \text{
        $\upiota$ non-decreasing and non-increasing in 
        $\mathcal{K}_\text{low}$ and $\mathcal{K}_\text{high}$,
      }
      \nonumber\\&
      \text{and
        $\upiota(\kappa)\geq\upiota(\kappa_{J+1-j_0})\,
        \forall \kappa \in \mathcal{K}_\text{mid}$
      }
    \biggr\}.
  \end{IEEEeqnarray}
\end{assumption}

If the basis-function expansion \eqref{eq:upiota} holds,
\eqref{eq:upiotaShape} reduces to
\begin{IEEEeqnarray}{rl}
  \label{eq:bcIShape}
  \mathcal{A}=\biggl\{
    \bcI\in\mathbb{R}_+^J
    \,\Big|\,
    \cI_1\leq\cI_2\leq\ldots&\leq\cI_{J+1-j_0},\;
    \cI_{j_0}\geq\ldots\geq\cI_{J-1}\geq\cI_{J},\;
    \nonumber\\&
    \text{ and } \cI_{j}\geq\cI_{J+1-j_0},\;
    \forall j \in [J+1-j_0,j_0]
  \biggr\}
\end{IEEEeqnarray}
see also Fig.~\ref{fig:spline}.  Here, the monotonic low- and high-$\kappa$ 
regions each contains $J-j_0$ knots in the B1-spline representation, 
whereas the central region contains $2j_0-J$ knots.

In practice, the X-ray spectrum $\iota(\varepsilon)$ starts at a lowest 
effective energy that can penetrate the object, vanishes at the tube 
voltage (the highest photon energy) and has a region in the center higher 
than the two ends, see Fig.~\ref{fig:triRelate}.  When the support of 
$\iota(\varepsilon)$ is free of $K$-edges, the mass attenuation coefficient 
$\kappa(\varepsilon)$ is monotonic, thus $\upiota(\kappa)$ (as a function 
of $\kappa$) has the same shape and trends as $\iota(\varepsilon)$ (as a 
function of $\varepsilon$) and justifies
Assumption~\ref{assumption:I}.  If a $K$-edge is present within the support 
of $\iota(\varepsilon)$, it is difficult to infer the shape of 
$\upiota(\kappa)$.  In most cases, Assumption~\ref{assumption:I} holds.

For the approximation of $\upiota(\kappa)$ using B1-spline basis expansion, 
as long as $[\kappa_0, \kappa_{J+1}]$ is sufficiently large to cover the 
range of $\kappa(\varepsilon)$ with 
$\varepsilon\in\supp(\iota(\varepsilon))$, we can always meet 
Assumption~\ref{assumption:I} by selecting $j_0$ appropriately.

Multiple different $(\balpha,\bcI)$ share the same noiseless output 
$\bcI^\tout(\balpha,\bcI)$ and thus the same \gls{NLL}, see 
Section~\ref{sec:ambiguity}.  In particular, equivalent $(\balpha,\bcI)$ 
can be constructed by left- or right-shifting the mass attenuation spectrum 
and properly rescaling it and the density map, see 
\eqref{eq:shiftambiguity}.  Selecting a fixed $j_0$ in \eqref{eq:j0} can 
exclude all these equivalent values but the one in which the 
mass-attenuation spectrum satisfies \eqref{eq:upiotaShape} and where the 
biconvexity of the \gls{NLL} can be established.

\begin{lem}
  \label{lemma:alphaconvexlognormal}
  Provided that Assumption~\ref{assumption:I} holds, the lognormal \gls{NLL} 
  $\cL_\upiota(\balpha)$ is a convex function of $\balpha$ over the following 
  region:
  \begin{subequations}
    \begin{IEEEeqnarray}{c}
      \label{eq:Lalphaconvexcond}
      \CBR{ \balpha \given{
        \upiota_\circ^\tL(\Phi \balpha) \succeq \E^{-U}\bcE,\; 
        \balpha\in\mathbb{R}_+^p
      }}
    \end{IEEEeqnarray}
    where
    \begin{equation}
      \label{eq:U}
      U \df \frac{2q^{j_0}}{(q^{j_0}-1)^2}.
    \end{equation}
  \end{subequations}
\end{lem}
\begin{IEEEproof}
  See Appendix~\ref{app:biConvex}.
\end{IEEEproof}

Upon taking the logarithm of both sides and rearranging, the condition in 
\eqref{eq:Lalphaconvexcond} corresponds to upper-bounding the residuals 
$\ln \cE_n - \ln \cI_n^\tout(\balpha,\bcI)$ by $U$.  The bound $U$ depends 
only on the common ratio $q$ and constant $j_0$ used to describe the 
constraint on $\upiota(\kappa)$ in Assumption~\ref{assumption:I}.  Note 
that Lemma~\ref{lemma:alphaconvexlognormal} does not assume a 
basis-function expansion of the mass-attenuation spectrum, only that it 
satisfies \eqref{eq:upiotaShape}.

If we wish $\upiota(\kappa)$ to cover the same range [i.e., 
\eqref{eq:qJcond} holds], then reducing $q$ needs to be accompanied by 
increasing $J$, which also leads to a larger $j_0$.  Indeed, $q^{j_0}$ in 
$U$ is the ratio of the point where $\upiota(\kappa)$ starts to be 
monotonically decreasing to the point where the support of 
$\upiota(\kappa)$ starts, see Fig.~\ref{fig:spline}.

\subsubsection{NLL of $\bcI$}
\label{sec:NLLI}
Fix $\balpha$ and define
\begin{subequations}
  \begin{IEEEeqnarray}{c}
    \label{eq:A}
    A= \bb_\circ^\tL( \Phi \balpha  ).
  \end{IEEEeqnarray}
  The \gls{NLL} of $\bcI$ for fixed $\balpha$ reduces to a Gaussian \gls{GLM} 
  for measurements $\ln_\circ(\bcE)$ with design matrix $A$ and exponential 
  link function:
  \begin{IEEEeqnarray}{c}
    \label{eq:lognormalNLLI}
    \cL_A(\bcI) =  \frac{1}{2}\normlr{
      \ln_\circ \bcE - \ln_\circ( A \bcI )
    }_2^2.
  \end{IEEEeqnarray}
\end{subequations}

In the following, we first provide the condition for the convexity of 
$\cL_A(\bcI)$:
\begin{lem}
  \label{lemma:iconvexlognormal}
  The \gls{NLL}  $\cL_A(\bcI)$ in \eqref{eq:lognormalNLLI} is a convex 
  function of $\bcI$ in the following region:
  \begin{equation}
    \label{eq:convexCondI}
    \CBR{\bcI\given{A\bcI\preceq\E\bcE}}.
  \end{equation}
  Upon taking the logarithm of both sides and rearranging, the condition in 
  \eqref{eq:convexCondI} corresponds to $\ln_\circ\bcI^\tout(\balpha,\bcI) 
  - \ln_\circ\bcE \preceq\bm{1}$.
\end{lem}
\begin{IEEEproof}
Since $\frac{\partial A \bcI }{\partial\bcI^T}=A$, we have
\begin{subequations}
  \begin{IEEEeqnarray}{rCl}
    \label{eq:gradientILognormal}
    \frac{\partial \cL_A(\bcI)}{\partial\bcI}
    &=&  A^T\diag^{-1}(A\bcI)
    \SBR{\ln_\circ(A \bcI) - \ln_\circ\bcE}
    \\
    \label{eq:hessianILognormal}
    \frac{\partial^2 \cL_A(\bcI)}{\partial\bcI\partial\bcI^T}
    &=&  A^T \diag^{-2}(A \bcI)
    \diagp{\bm{1}-\ln_\circ(A \bcI) + \ln_\circ\bcE} A
  \end{IEEEeqnarray}
\end{subequations}
According to \eqref{eq:convexCondI}, 
$\diag\PARENS{\bm{1}-\ln_\circ(A\bcI)+\ln_\circ\bcE}\succeq \bm{0}$, the 
Hessian \eqref{eq:hessianILognormal} is positive semidefinite, and 
$\cL_A(\bcI)$ is convex.
\end{IEEEproof}

From the Hessian expression in \eqref{eq:hessianILognormal},  we conclude 
that $\cL_A(\bcI)$ in \eqref{eq:lognormalNLLI} is strongly convex if the 
condition \eqref{eq:convexCondI} holds with strict inequality and design 
matrix $A$ has full rank.

Combining the convexity results in Lemmas~\ref{lemma:alphaconvexlognormal} 
and \ref{lemma:iconvexlognormal} yields the biconvexity region for the 
lognormal \gls{NLL} $\cL\PARENS{\balpha,\bcI}$ in \eqref{eq:lognormalNLL}, 
see Section~\ref{sec:biconvexity}.

\subsection{Poisson measurement model}
\label{sec:Poisson}

For an $N \times 1$ vector $\bcE$ of independent Poisson measurements, the 
\gls{NLL} in the form of generalized Kullback-Leibler divergence 
\cite{ZanniBertero2014} is [see also \eqref{eq:IinIoutparI}]
\begin{IEEEeqnarray}{rCl}
  \label{eq:poissNLL}
  \cL(\balpha,\bcI)=\bm{1}^T\SBR{\bcI^\tout(\balpha,\bcI)-\bcE}
  -\bcE^T\SBR{\ln_\circ \bcI^\tout(\balpha,\bcI) - \ln_\circ\bcE}.
\end{IEEEeqnarray}

In the following, we express \eqref{eq:poissNLL} as a function of $\balpha$ 
with $\bcI$ fixed (Section~\ref{sec:NLLalphaPois}) and vice versa (Section 
\ref{sec:NLLIPois}), which will be used to describe our estimation 
algorithm in Section~\ref{sec:estimation}.

\subsubsection{NLL of $\balpha$}
\label{sec:NLLalphaPois}
Recall \eqref{eq:upiota} and write the \gls{NLL} of $\balpha$ for fixed 
$\bcI$ as 
\begin{IEEEeqnarray}{rCl}
  \label{eq:poissonNLLalpha}
  \cL_\upiota\PARENS{\balpha} &=& 
  \bm{1}^T\SBRbig{\upiota_\circ^\tL(\Phi\balpha)-\bcE}
  -\bcE^T\SBRbig{\ln_\circ \upiota_\circ^\tL(\Phi\balpha)-\ln_\circ\bcE}
\end{IEEEeqnarray}
which corresponds to the Poisson \gls{GLM} with design matrix $\Phi$ and 
link function equal to the inverse of  $\upiota^\tL(\cdot)$.

\begin{lem}
  \label{lemma:alphaconvexPoisson}
  Provided that Assumption~\ref{assumption:I} holds, the Poisson \gls{NLL} 
  $\cL_\upiota(\balpha)$ is
  a convex function of $\balpha$ over the following region:
  \begin{subequations}
    \begin{equation}
      \mathcal{P}=\CBR{\balpha
        \given{
          \upiota_\circ^\tL(\Phi\balpha) \succeq (1-V)\bcE,\; \balpha\in 
          \mathbb{R}_+^p
        }
      }
      \label{eq:poissonConvCondalpha}
    \end{equation}
    where
    \begin{equation}
      \label{eq:V}
      V \df \frac{2q^{j_0}}{q^{2j_0}+1}.
    \end{equation}
  \end{subequations}
\end{lem}
\begin{IEEEproof}
  See Appendix~\ref{app:biConvex}.
\end{IEEEproof}
Note that the region in \eqref{eq:poissonConvCondalpha} is only sufficient 
for convexity and that Lemma~\ref{lemma:alphaconvexPoisson} does not assume 
a basis-function expansion of the mass-attenuation spectrum, only that it 
satisfies \eqref{eq:upiotaShape}.

Similar to $U$, $1-V$, the lower bound of $\cI_n^\tout(\balpha,\bcI) \big/ 
\cE_n$ for all $n$, is also a function of $q^{j_0}$, which is determined by 
the shape of $\upiota(\kappa)$.

\subsubsection{NLL of $\bcI$}
\label{sec:NLLIPois}
For fixed $\balpha$,
the \gls{NLL} of $\bcI$ reduces to [see \eqref{eq:A}]
  \begin{IEEEeqnarray}{rCl}
  \label{eq:poissonNLLI}
  \cL_A\PARENS{\bcI} &=& \bm{1}^T\PARENS{A\bcI-\bcE}
  -\bcE^T\SBR{\ln_\circ\PARENS{A\bcI}-\ln_\circ\bcE}
\end{IEEEeqnarray}
which corresponds to the Poisson \gls{GLM} with design matrix $A$  in 
\eqref{eq:A} and identity link. 

\begin{lem}
  \label{lemma:IconvexPoisson}
  The \gls{NLL}  $\cL_A(\bcI)$ in \eqref{eq:poissonNLLI} is a convex function 
  of $\bcI$ for all $\bcI \in \mathbb{R}_+^J$.
\end{lem}

\begin{IEEEproof}
  The gradient and Hessian of the \gls{NLL} in \eqref{eq:poissonNLLI} are
  \begin{subequations}
    \begin{IEEEeqnarray}{rCl}
      \frac{\partial \cL_A(\bcI)}{\partial\bcI}
      &=&  A^T
      \SBRbig{\bm{1}-\diag^{-1}(A\bcI)\bcE}
      \\
      \label{eq:hessianIPoisson}
      \frac{\partial^2 \cL_A(\bcI)}{\partial\bcI\partial\bcI^T}
      &=&  A^T \diag(\bcE) \diag^{-2}(A\bcI)A
    \end{IEEEeqnarray}
  \end{subequations}
  where the Hessian matrix \eqref{eq:hessianIPoisson} is positive 
  semidefinite.  Thus, $\cL_A(\bcI)$ in \eqref{eq:poissonNLLI} is convex on 
  $\mathbb{R}_+^J$.
\end{IEEEproof}

From the Hessian expression in \eqref{eq:hessianIPoisson},  we conclude 
that  $\cL_A(\bcI)$ in \eqref{eq:poissonNLLI} is strongly convex if the 
design matrix $A$ has full rank.

Combining the convexity results in Lemmas~\ref{lemma:alphaconvexPoisson} 
and \ref{lemma:IconvexPoisson} yields the biconvexity region for the 
\gls{NLL} $\cL\PARENS{\balpha,\bcI}$ in \eqref{eq:poissNLL}, see the 
following section.

\subsection{Biconvexity of the \glspl{NLL}}
\label{sec:biconvexity}

\begin{thm}
  \label{th:biConvex}
  Suppose that Assumption~\ref{assumption:I} in \eqref{eq:bcIShape} holds.  
  Then,
  \begin{subequations}
    \begin{enumerate}[label=\Roman*), ref=\Roman*)]
      \item  \label{th:biConvexlognormal}
        the lognormal \gls{NLL} function \eqref{eq:lognormalNLL} is 
        biconvex \cite{Gorski2007Biconvex} with respect to $\balpha$ and 
        $\bcI$ in the region
        \begin{equation}
          \label{eq:lognormalBiConvCond}
          \mathcal{G}=\CBR{ \PARENS{\balpha,\bcI} \given{
            \E^{-U}\bcE\preceq\bcI^\tout(\balpha,\bcI)\preceq\E\bcE,\; 
            \bcI\in\mathcal{A},\; \balpha\in\mathbb{R}_+^p
          }}
        \end{equation}
        which bounds the elements of the residual vector $\ln_\circ \bcE - 
        \ln_\circ \bcI^\tout(\balpha,\bcI)$ [whose squared $\ell_2$-norm we 
        wish to minimize, see \eqref{eq:lognormalNLL}] between $-1$ and 
        $U$, and
      \item \label{th:biConvexPoisson}
        the Poisson \gls{NLL} \eqref{eq:poissNLL} is biconvex with respect 
        to $\balpha$ and $\bcI$ in the following set:
        \begin{equation}
          \mathcal{P}=\CBR{(\balpha,\bcI)
            \given{
              \bcI^\tout(\balpha,\bcI)\succeq (1-V)\bcE,\; 
              \bcI\in\mathcal{A},\; \balpha\in \mathbb{R}_+^p
            }
          }
          \label{eq:poissonBiConvCond}
        \end{equation}
        which bounds $\cI_n^\tout(\alpha,\bcI) \big/ \cE_n$
        from below by $1-V$ for all $n$, see also \eqref{eq:V}.
    \end{enumerate}
  \end{subequations}
\end{thm}
\begin{IEEEproof}
We first show the convexity of the regions defined in 
\eqref{eq:lognormalBiConvCond} and \eqref{eq:poissonBiConvCond} with 
respect to each variable ($\balpha$ and $\bcI$) with the other fixed.  We 
then show the convexity of the \gls{NLL} functions \eqref{eq:lognormalNLL} 
and \eqref{eq:poissNLL} for each variable.
  
The region $\mathcal{A}$ in \eqref{eq:bcIShape} is a subspace, thus a 
convex set.  Since $\bcI^\tout$ in \eqref{eq:IinIoutparI} is a linear 
function of $\bcI$, the inequalities comparing $\bcI^\tout$ to constants 
specify a convex set.  Therefore, both $\mathcal{G}_\balpha \df \{\bcI 
  \given{ (\balpha,\bcI)\in\mathcal{G}}\}$ and $\mathcal{P}_\balpha=\{\bcI 
    \,|\, {(\balpha,\bcI)\in\mathcal{P}}\}$ are convex for fixed  
    $\balpha\in\mathbb{R}_+^p$ because both are intersections between the 
    subspace $\mathcal{A}$ and a convex set via $\bcI^\tout$.  Since 
    $b_j(\kappa)\geq0$,  $\PARENSbig{b_j^\tL(s)}_{j=1}^J = 
    \int_{\kappa_{j-1}}^{\kappa_{j+1}}b_j(\kappa)\E^{-s\kappa}\dif\kappa$ 
    are  decreasing functions of $s$, which, together with the fact that 
    $\bcI\succeq \bm{0}$, implies that $\bb^\tL(\bs)\bcI$ is a decreasing 
    function of $\bs$.  Since the linear transform $\Phi\balpha$ preserves 
    convexity, both $\mathcal{G}_\bcI=\{\balpha  \,|\, 
      (\balpha,\bcI)\in\mathcal{G}\}$ and $\mathcal{P}_\bcI=\{\balpha \,|\, 
        (\balpha,\bcI)\in\mathcal{P}\}$ are convex with respect to 
        $\balpha$ for fixed $\bcI \in \mathcal{A}$.  Therefore, 
        $\mathcal{G}$ and $\mathcal{P}$ are biconvex with respect to $\bcI$ 
        and $\balpha$.

Observe that $\mathcal{G}$ in \eqref{eq:lognormalBiConvCond} is the 
intersection of the regions specified by Assumption~\ref{assumption:I} and 
Lemmas~\ref{lemma:alphaconvexlognormal} and \ref{lemma:iconvexlognormal}.  
Thus, within $\mathcal{G}$, the lognormal \gls{NLL} \eqref{eq:lognormalNLL} 
is a convex function of $\balpha$ (with fixed $\bcI$) and $\bcI$ (with 
fixed $\balpha$), respectively.  Similarly, $\mathcal{P}$ in 
\eqref{eq:poissonBiConvCond} is the intersection of the regions specified 
by Assumption~\ref{assumption:I} and Lemmas~\ref{lemma:alphaconvexPoisson} 
and \ref{lemma:IconvexPoisson}.  Thus, within $\mathcal{P}$, the Poisson 
\gls{NLL} \eqref{eq:poissNLL} is a convex function of $\balpha$ (with fixed 
$\bcI$) and $\bcI$ (with fixed $\balpha$), respectively. 

By combining the above region and function convexity results, we conclude 
that \eqref{eq:lognormalNLL} and \eqref{eq:poissNLL} are biconvex within 
$\mathcal{G}$ and $\mathcal{P}$, respectively.
\end{IEEEproof}

\section{Density-Map and Mass-Attenuation Spectrum Estimation}
\label{sec:estimation}

\subsection{Penalized \gls{NLL} objective function and its properties}
\label{sec:obj}

Our goal is to compute
penalized maximum-likelihood estimates of the density-map and 
mass-attenuation spectrum parameters $\PARENS{ \balpha,  \bcI }$ by solving  
the following minimization problem:  \label{eq:sparseestimationproblem0}
\begin{subequations}
  \begin{IEEEeqnarray}{c}
    \label{eq:constropt}
    \min_{ \balpha, \bcI}
    f(\balpha,\bcI)
  \end{IEEEeqnarray}
  where
  \begin{IEEEeqnarray}{c}
    f(\balpha,\bcI)=\cL(\balpha,\bcI)+ur(\balpha)+\mathbb{I}_\nonneg(\bcI)
    \label{eq:obj}
  \end{IEEEeqnarray}
\end{subequations}
is the penalized \gls{NLL} objective function, $u>0$ is a scalar tuning 
constant, the density-map regularization term $r(\balpha)$ enforces 
nonnegativity and sparsity of the signal $\balpha$ in an appropriate 
transform [e.g., \gls{TV} or \gls{DWT}] domain, and the third summand 
enforces the nonnegativity of the mass-attenuation spectrum parameters 
$\bcI$, see \eqref{eq:Inonneg}.  We consider the following two 
$r(\balpha)$:
\begin{subequations}
  \label{eq:r}
  \begin{IEEEeqnarray}{rCl}
    \label{eq:rTV}
    r(\balpha)&=& \mathlarger{\sum}_{i=1}^p 
    \sqrt{\sum_{j\in\mathcal{N}_i}(\alpha_i-\alpha_j)^2}
    + \mathbb{I}_\nonneg(\balpha)\\
    \label{eq:rWV}
    r(\balpha)&=& \bigl\| \Psi^T \balpha \bigr\|_1 + 
    \mathbb{I}_\nonneg(\balpha)
  \end{IEEEeqnarray}
\end{subequations}
where the second summands enforce the nonnegativity of $\balpha$.  The 
first summand  in \eqref{eq:rTV}  is an isotropic \gls{TV}-domain sparsity 
term \cite{RudinOsherFatemi1992,Beck2009TV}  and $\mathcal{N}_i$ is index 
set of neighbors of the $i$th element of $\balpha$, where  elements of 
$\balpha$ are arranged to form a 2-dimensional image.  Each set 
$\mathcal{N}_i$ consists of two pixels at most, with one on the top and the 
other on the right of the $i$th pixel, if possible \cite{Beck2009TV}.  The 
first summand  in \eqref{eq:rWV}  imposes sparsity of transform-domain 
density-map coefficients $\Psi^T \balpha$, where the sparsifying dictionary 
$\Psi \in \mathbb{R}^{p \times p'}$ ($p \leq p'$) is assumed to satisfy the 
orthogonality condition \begin{IEEEeqnarray}{c}
  \label{eq:Psiorth} \Psi\Psi^T=I_p.
\end{IEEEeqnarray}
In \cite{gdqnde14}, we used the sparsity regularization term \eqref{eq:rWV}
with \eqref{eq:Psiorth} and the lognormal \gls{NLL} in 
\eqref{eq:lognormalNLL}.

Since both $r(\balpha)$ in \eqref{eq:rTV} and \eqref{eq:rWV} are convex for 
any $\balpha$, and $\mathbb{I}_\nonneg(\bcI)$ in \eqref{eq:obj} is convex 
for all $\bcI$, the following holds.
\begin{cor}
  \label{fbiconvex}
  The objective $f(\balpha,\bcI)$ in \eqref{eq:obj} is biconvex with 
  respect to $\balpha$ and $\bcI$ under the conditions specified by 
  Theorem~\ref{th:biConvex}.
\end{cor}

Although the \gls{NLL} may have multiple local minima of the form 
$q^n\what{\balpha}$ (see Section~\ref{sec:ambiguity}), those with large $n$ 
can be eliminated by the regularization penalty.  To make it clear, let us 
first see the impact the ambiguity has on the scaling of the first 
derivative of the objective function $f(\balpha,\bcI)$.  From 
\eqref{eq:shiftambiguity}, we conclude that $\bz_0=(\balpha, 
[0,\cI_2,\dotsc,\cI_J]^T)$ and $\bz_1=(q \balpha, q 
[\cI_2,\dotsc,\cI_J,0]^T)$ have the same $\bcI^\tout(\balpha,\bcI)$ and 
thus the same \gls{NLL}.  The first derivative of 
$\cL(\balpha,\bcI)=\cL(\bz)$ over $\balpha$ at $\bz_1$ is $1/q$ times that 
at $\bz_0$.  Meanwhile, the subgradients of the regularization term at 
$\bz_0$ and $\bz_1$ with respect to $\balpha$ are the same. Hence, with the 
increase of $n$, the slope of $f(\balpha,\bcI)$ is dominated by the penalty 
term.  This is also experimentally confirmed: we see the initialization 
$q^n \step{\balpha}{0}$ with large $n$ being reduced as the iteration 
proceeds.

We now establish that the objective function \eqref{eq:obj} satisfies the 
\gls{KUL} property \cite{AttouchBolte2010}, which is important for 
establishing local convergence of block-coordinate schemes in biconvex 
optimization problems.  
\begin{thm}[\gls{KUL} Property]
  \label{th:kl}
  The objective function $f(\balpha,\bcI)$ satisfies the \gls{KUL} property 
  in any compact subset $\mathbb{C}\subseteq\dom(f)$. Here, the \gls{NLL}
  $\cL(\balpha,\bcI)$ is either lognormal or Poisson in 
  \eqref{eq:lognormalNLL} or \eqref{eq:poissNLL}, respectively,
  and the penalty $r(\balpha)$ in
  \eqref{eq:obj} is given in \eqref{eq:rTV}
  or \eqref{eq:rWV}.
\end{thm}
\begin{IEEEproof}
  See Appendix~\ref{app:KL_PropertyProofConv}. 
\end{IEEEproof}

Note that the domain of $f$ requires positive measurements such that  
$\bcI^\tout(\balpha,\bcI) \succ\bm{0}$, which excludes the case 
$\bcI=\bm{0}$ when no incident X-ray is applied, see also 
\eqref{eq:IinIoutparI}.
The \emph{compact set} assumption keeps the distance between $\bcI$ and 
$\bm{0}$ from going to zero.

\subsection{Minimization algorithm}

The parameters that we wish to estimate are naturally divided into two 
blocks, $\balpha$ and $\bcI$.  The large size of $\balpha$ prohibits 
effective second-order methods under the sparsity regularization, whereas 
$\bcI$ has much smaller size and only nonnegative constraints, thus 
allowing for more sophisticated solvers, such as the quasi-Newton 
\gls{BFGS} approach \cite[Sec.~4.3.3.4]{Thisted} that we adopt here.  In 
addition, the scaling difference between $\balpha$ and $\bcI$ can be 
significant so that the joint gradient method for $\balpha$ and $\bcI$ 
together would converge slowly.  Therefore, we adopt a block 
coordinate-descent algorithm to minimize $f(\balpha,\bcI)$ in 
\eqref{eq:obj}, where the \gls{NPG} \cite{gdqnde14} and \gls{LBFGSB} 
\cite{Byrd1995LBFGS} methods are employed to update estimates of the 
density-map and mass-attenuation spectrum parameters, respectively.
The choice of block coordinate-descent optimization is also motivated by 
the related \gls{ACS} and block coordinate-descent schemes in 
\cite{Gorski2007Biconvex} and \cite{XuYin2013}, respectively, both with 
convergence guarantees under certain conditions.

We minimize the objective function \eqref{eq:obj} by alternatively updating 
$\balpha$ and $\bcI$ using \ref{alphaStep} and \ref{bcIStep}, respectively, 
where Iteration~$i$ proceeds as follows:
\begin{enumerate}[leftmargin=0.75in, label= Step \arabic*)]
  \item (NPG) \label{alphaStep}
    Set the mass-attenuation spectrum $\upiota(\kappa) = \bb(\kappa) 
    \step{\bcI}{i-1}$, treat it as known\footnote{
      This selection corresponds to $\cL_\upiota(\balpha)=  
      \cL\PARENSbig{\balpha,\step{\bcI}{i-1}}$, see also 
      \eqref{eq:lognormalNLLalpha} and \eqref{eq:poissonNLLalpha}.
    }, and descend the regularized \gls{NLL} function $f(\balpha, 
    \step{\bcI}{i-1})=\cL_\upiota(\balpha) + u r(\balpha)$ by applying an 
    \emph{\gls{NPG} step} for $\balpha$, which yields $\step{\balpha}{i}$:
    \begin{subequations}
      \label{eq:NPGstepalpha}
      \begin{IEEEeqnarray}{rCl}
        \label{eq:thetaupdate}
        \theta^{(i)} &=& \frac{1}{2}
        \SBR{ 1 + \sqrt{ 1 + 4 \bigl(\theta^{(i-1)}\bigr)^2 } }
        \\
        \label{eq:nesterov}
        \wbalpha^{(i)} &=& \balpha^{(i-1)}
        + \frac{\theta^{(i-1)}-1}{\theta^{(i)}}
        \bigl( \balpha^{(i-1)} - \balpha^{(i-2)} \bigr)\\
        \label{eq:objAlphaStep}
        \balpha^{(i)}&=&\arg\min_\balpha \frac{1}{2\beta^{(i)}}
        \bigl\| \balpha-\wbalpha^{(i)}
        +\beta^{(i)}\nabla \cL_\upiota\bigl( \wbalpha^{(i)}\bigr)
        \bigr\|_2^2 + ur(\balpha)
      \end{IEEEeqnarray}
      where $\beta^{(i)}>0$ is an adaptive step size chosen to satisfy the 
      majorization condition:
      \begin{IEEEeqnarray}{rCl}
        \label{eq:stepCond}
        \cL_\upiota\bigl(\balpha^{(i)}\bigr)
        \leq \cL_\upiota\bigl(\wbalpha^{(i)}\bigr)
        &+& \bigl(\balpha^{(i)}-\wbalpha^{(i)}\bigr)^T
        \nabla \cL_\upiota\bigl( \wbalpha^{(i)}\bigr)
        + \frac{1}{2\beta^{(i)}}
        \bigl\| \balpha^{(i)}-\wbalpha^{(i)}\bigr\|_2^2
      \end{IEEEeqnarray}
    \end{subequations}
    using a simple adaptation scheme that aims at keeping $\step{\beta}{i}$ 
    as large as possible, as long as it satisfies \eqref{eq:stepCond}, see 
    Section~\ref{sec:alphaStep} for details; we also apply the ``function 
    restart'' \cite{ODonoghue2013} to restore the monotonicity and improve 
    convergence of  \gls{NPG} steps.
  \item (BFGS) \label{bcIStep}
    Set the design matrix $A= \bb_\circ^\tL\bigl( \Phi \step{\balpha}{i} 
    \bigr)$, treat it as known\footnote{
      This selection corresponds to 
      $\cL_A(\bcI)=\cL\PARENSbig{\balpha^{(i)},\bcI}$, see also 
      \eqref{eq:lognormalNLLI} and \eqref{eq:poissonNLLI}.
    }, and minimize the regularized \gls{NLL} function 
    $f\bigl(\step{\balpha}{i},\bcI\bigr)$ with respect to $\bcI$, i.e.,  
    \begin{equation}
      \label{eq:bcIStep}
      \bcI^{(i)} = \arg\min_{\bcI\succeq\bm{0}} \cL_A(\bcI)
    \end{equation}
    using the inner \gls{LBFGSB} iteration, initialized by 
    $\step{\bcI}{i-1}$.
\end{enumerate}
We refer to this iteration as the \emph{NPG-BFGS algorithm}.  

The above iteration is the first physical-model based image reconstruction 
method (in addition to our preliminary work in \cite{GuDogandzic2013}) for 
simultaneous \emph{blind} (assuming unknown incident X-ray spectrum and 
unknown materials) sparse image reconstruction from polychromatic 
measurements.  In \cite{GuDogandzic2013}, we applied a piecewise-constant 
(B0 spline) expansion of the mass attenuation spectrum, approximated 
Laplace integrals with Riemann sums, used a smooth approximation of the 
nonnegativity penalties in \eqref{eq:r} instead of exact, and did not 
employ signal-sparsity regularization.

The optimization task in \eqref{eq:objAlphaStep} is a proximal step:
\begin{subequations}
  \begin{IEEEeqnarray}{rCl}
    \label{eq:proxgradstep}
    \balpha^{(i)} &=& \proxp{\beta^{(i)} u r} { \wbalpha^{(i)} -
    \beta^{(i)} \nabla\cL_\upiota\bigl( \wbalpha^{(i)} \bigr)}
  \end{IEEEeqnarray}
  where the proximal operator for scaled (by $\lambda>0$) regularization term 
  \eqref{eq:r} is defined as \cite{Parikh2013Proximal}
  \begin{equation}
    \proxp{\lambda r}{\ba} = \arg \min_{\balpha}
    \frac{1}{2}\norm{\balpha-\ba}_2^2 + \lambda r(\balpha).
    \label{eq:proximallambda}
  \end{equation}
\end{subequations}
If we \emph{do not} apply the Nesterov's acceleration 
\eqref{eq:thetaupdate}--\eqref{eq:nesterov} and use only the \gls{PG} step 
\eqref{eq:objAlphaStep} to update the density-map iterates $\balpha$, i.e., 
\begin{IEEEeqnarray}{c}
  \label{eq:nomomentum}
  \wbalpha^{(i)}=\balpha^{(i-1)}
  \end{IEEEeqnarray}
then the corresponding iteration is the \emph{PG-BFGS algorithm}, see also 
the numerical examples and Fig.~\ref{fig:castingTrace}.  We show that our 
PG-BFGS algorithm is monotonic (Section~\ref{sec:alphaStep}) and converges 
to one of the critical points of the objective $f(\balpha,\bcI)$ 
(Section~\ref{sec:convergence}).  Unfortunately, such theoretical 
convergence properties do not carry over to NPG-BFGS iteration, yet this 
iteration empirically outperforms the PG-BFGS method, see 
Section~\ref{sec:numex}.

If the mass-attenuation spectrum $\upiota(\kappa)$ is known and we iterate 
\ref{alphaStep} only to estimate the density map $\balpha$, we refer to 
this iteration as the \emph{\gls{NPG} algorithm} for $\balpha$.

\ref{alphaStep} employs only one \gls{NPG} step, where an inner iteration 
is needed to compute the proximal step \eqref{eq:objAlphaStep}, whereas 
\ref{bcIStep} employs a full \gls{LBFGSB} iteration.  Hence, both 
\ref{alphaStep} and \ref{bcIStep} require inner iterations.  To solve 
\eqref{eq:proximallambda} for the two regularizations, we adopt 
\begin{itemize}
  \item \gls{TV}-based denoising method  in \cite[Sec.~IV]{Beck2009TV} for 
    the \gls{TV} regularization \eqref{eq:rTV} and
  \item the \gls{ADMM} method for the \gls{DWT}-based regularization 
    \eqref{eq:rWV}, described in the following section.
  \end{itemize}

\subsubsection{Proximal Mapping using ADMM for \gls{DWT}-based 
regularization}
\label{sec:proximalmapping}

We now present an \gls{ADMM} scheme for computing the proximal operator in 
\eqref{eq:proximallambda} for the \gls{DWT}-based regularization 
\eqref{eq:rWV} (see 
\cite[Appendix~\ref{nesterovreport-app:ADMM}]{NesterovTechReport}):
\begin{subequations}
  \label{eq:ADMM}
  \begin{IEEEeqnarray}{rCl}
    \label{eq:truncate}
    \step{\balpha}{k+1} &=&
    \frac{1}{1+\rho}\PARENS{
      \ba+\rho\Psi\PARENS{ \step{\bs}{k}+\step{\bupsilon}{k} }
    }_+ \\
    \label{eq:sUpdate}
    \step{\bs}{k+1} &=& \softthrp{\frac{\lambda}{\rho}}
    {\Psi^T\step{\balpha}{k+1}-\step{\bupsilon}{k}}\\
    \label{eq:dualUpdate}
    \step{\bupsilon}{k+1} &=&
    \step{\bupsilon}{k}+\step{\bs}{k+1}-\Psi^T\step{\balpha}{k+1}
  \end{IEEEeqnarray}
\end{subequations}
where $k$ is the inner-iteration index and $\rho>0$ is a step-size 
parameter, usually set to 1 \cite[Sec.~11]{Boyd2011ADMM}. We obtain 
\eqref{eq:ADMM}  by decomposing the proximal objective function 
\eqref{eq:proximallambda} into the sum of 
$\frac{1}{2}\normlr{\balpha-\ba}_2^2 + \mathbb{I}_\nonneg(\balpha)$ and 
$\lambda \norm{\Psi^T\balpha}_1$.  The above algorithm is initialized by
\begin{IEEEeqnarray}{c}
  \label{eq:ADMMinit}
  \step{\bs}{0}=\Psi^T\ba
  \qquad \text{and} \qquad
  \step{\bupsilon}{0}=\bm{0}_{p' \times 1}
\end{IEEEeqnarray}
which is equivalent to projecting the noisy signal $\ba$
onto the nonnegative signal space $\balpha^{(1)}=(\ba)_+$.
Note that the signal estimate $\PARENSbig{\Psi\step{\bs}{j}}_+$   is within 
the signal space and yields finite objective function 
$\frac{1}{2}\norm{\balpha-\ba}_2^2 + \lambda r(\balpha)$ that we wish to 
minimize; we use it to obtain the final signal estimate upon convergence of 
the \gls{ADMM} iteration.

In the following, we describe the convergence criteria for  the outer and 
inner iterations (Section~\ref{sec:convCrit}), adaptive step-size selection 
(Section~\ref{sec:alphaStep}), algorithm initialization 
(Section~\ref{sec:init}), and convergence analysis 
(Section~\ref{sec:convergence}).

\subsubsection{Convergence criteria}
\label{sec:convCrit}

Define the measures of change of the density map $\balpha$ and the 
\gls{NLL}
\begin{subequations}
\begin{IEEEeqnarray}{rCl}
  \label{eq:delta}
  \step{\delta}{i} &=& \bigl\lVert \step{\balpha}{i}-\step{\balpha}{i-1} 
  \bigr\rVert_2
  \\
  \label{eq:deltaL}
  \step{\delta_\cL}{i} &=& \abslr{
    \cL\bigl(\balpha^{(i)},\bcI^{(i-1)}\bigr)
    -\cL\bigl(\balpha^{(i-1)},\bcI^{(i-1)}\bigr)
  }
\end{IEEEeqnarray}
\end{subequations}
upon completion of \ref{alphaStep} in Iteration~$i$.
We run the outer iteration between \ref{alphaStep} and \ref{bcIStep} until
the relative distance of consecutive iterates of the density map $\balpha$ 
does not change significantly:
\begin{IEEEeqnarray}{c}
  \label{eq:convConNPGAS}
  \step{\delta}{i}   < \epsilon 
  \bigl\lVert\step{\balpha}{i}\bigr\rVert_2\end{IEEEeqnarray}
where $\epsilon > 0$ is the convergence threshold.

\textbf{Proximal step \eqref{eq:objAlphaStep}.}
The convergence criteria for the inner \gls{TV}-denoising and \gls{ADMM}  
iterations used to solve \eqref{eq:objAlphaStep} in Iteration~$i$ are based 
on the relative change $\step{\delta}{i-1}$:
\begin{subequations}
  \begin{IEEEeqnarray}{rCl}
    \label{eq:innerrit}
    \bigl\| \step{\balpha}{i,k} - \step{\balpha}{i,k-1}  \bigr\|_2
    &<& \eta_\balpha \step{\delta}{i-1}
    \\
    \label{eq:admmCrit}
    \max\CBR{
      \bigl\| \step{\bs}{i,k}-\Psi^T\step{\balpha}{i,k} \bigr\|_2,
      \bigl\| \step{\bs}{i,k}-\step{\bs}{i,k-1} \bigr\|_2
    } &<& \eta_\balpha \step{\delta}{i-1}
  \end{IEEEeqnarray}
\end{subequations}
where $k$ are the inner-iteration indices and the convergence tuning 
constant $\eta_\balpha \in(0,1)$ is chosen to trade off the accuracy and 
speed of the inner iterations and provide sufficiently accurate solutions 
to \eqref{eq:objAlphaStep}.  Note that $\bs^{(i,k)}$ in \eqref{eq:admmCrit} 
is the dual variable in \gls{ADMM} that converges to 
$\Psi^T\balpha^{(i,k)}$, and the criterion is on both the \emph{primal 
residual} $\bigl\| \step{\bs}{i,k}-\Psi^T\step{\balpha}{i,k} \bigr\|_2$ and 
the \emph{dual residual} $\bigl\| \step{\bs}{i,k}-\step{\bs}{i,k-1} 
\bigr\|_2$ \cite[Sec.~3.3]{Boyd2011ADMM}.

\textbf{BFGS step \eqref{eq:bcIStep}.}
For \ref{bcIStep} in Iteration~$i$, we set the convergence criterion as
\begin{equation}
  \bigl|\cL_A(\bcI^{(i,k)})-\cL_A(\bcI^{(i,k-1)})\bigr|
  <
  \eta_{\bcI} \step{\delta}{i}_\cL
  \label{eq:npgIcrit}
\end{equation}
where the tuning constant $\eta_{\bcI}\in(0,1)$ trades off the accuracy and 
speed of the estimation of $\bcI$. Since our main goal is to estimate the 
density map $\balpha$, the benefit to \ref{alphaStep} provided by 
\ref{bcIStep} through the minimization of $\cL_A(\bcI)$ in 
\eqref{eq:bcIStep} is more important than the reconstruction of $\bcI$ 
itself; furthermore, the estimation of $\bcI$ is sensitive to the rank of 
$A$.  Hence, we select $\step{\delta_\cL}{i}$ as a convergence metric for 
the inner iteration for $\bcI$ in \ref{bcIStep} of Iteration~$i$.

We safeguard both the proximal and \gls{BFGS} steps by setting the maximum 
number of iteration limit $n_\text{sub}$.  In summary, the outer and inner 
convergence tuning constants are $(\epsilon, \eta_\balpha, \eta_\bcI, 
n_\text{sub})$. The default values of these tuning constants are 
\begin{IEEEeqnarray}{c}
  \label{eq:convergenceconstants}
  \PARENS{ \epsilon, \eta_\balpha, \eta_\bcI, n_\text{sub} } = \PARENS{ 
    10^{-6}, 10^{-3}, 10^{-2}, 20}
\end{IEEEeqnarray}
adopted in all numerical examples in this paper.

\subsubsection{Adaptive step-size selection in \ref{alphaStep}}
\label{sec:alphaStep}
We adopt an adaptive step size selection strategy to accommodate the 
scenarios when the local Lipschitz constant varies as the algorithm 
evolves.

We select the step size $\beta^{(i)}$ to satisfy the majorization condition 
\eqref{eq:stepCond} using the following adaptation scheme 
\cite{NesterovTechReport,gdasil14}: 
\begin{enumerate}[label=\alph*)]
  \item In Iteration $i$: if there has been no step size reductions for 
    $\mathbbm{n}$ consecutive iterations, i.e., $\beta^{(i-1)} = 
    \beta^{(i-2)} = \dotsb = \beta^{(i-\mathbbm{n}-1)}$,  start with a 
    larger step size $\beta^{(i)}=\beta^{(i-1)}/\xi$, where $\xi \in (0,1)$ 
    is a \emph{step-size adaptation parameter}; otherwise start with 
    $\beta^{(i)}=\beta^{(i-1)}$;
  \item \emph{Backtrack}
    using the same multiplicative scaling constant $\xi$, with goal to find 
    the largest $\beta^{(i)}$ that satisfies 
    \eqref{eq:stepCond}.
\end{enumerate}
We select the initial step size $\beta^{(0)}$ using the Barzilai-Borwein 
method \cite{BarzilaiBorwein1988}.

\subsubsection{Initialization}
\label{sec:init}

Initialize the density-map and mass-attenuation coefficient vector iterates 
as
\begin{subequations}
  \begin{IEEEeqnarray}{rCl}
    \label{eq:initialrestart}
    \balpha^{(0)} &=& \balpha^{(-1)}= \what{\balpha}_\text{FBP},
    \qquad
    \theta^{(0)}=0\\
    \bcI^{(0)} &=& \frac{\max_n\cE_n}{b_{ \lceil (J+1)/2 \rceil
    }^\tL(0)}  \bm{e}_{ \lceil (J+1)/2 \rceil
    }  
    \label{eq:initI}
  \end{IEEEeqnarray}
\end{subequations}
where $\what{\balpha}_\text{FBP}$ is the standard \gls{FBP} reconstruction 
\cite[Ch.~3]{Kak1988CT}, \cite[Sec.~3.5]{Hsieh2009CT}.  We select the 
initial step size $\beta^{(0)}$ using the Barzilai-Borwein method 
\cite{BarzilaiBorwein1988}.  Plugging the initialization \eqref{eq:initI} 
into \eqref{eq:IinIoutparI} yields $\cI^\tin\bigl(\bcI^{(0)}\bigr)=  
\max_n\cE_n$  and the initial estimate 
$\bigl(\balpha^{(0)},\bcI^{(0)}\bigr)$ corresponds approximately to a 
monochromatic X-ray model; more precisely, it is a polychromatic X-ray 
model with a narrow mass attenuation spectrum proportional to $b_{\lceil 
(J+1)/2 \rceil}(\kappa)$.  It is also desirable to have the main lobe of 
the estimated spectrum at the center, which is why the nonzero element of 
$\bcI^{(0)}$ is placed in the middle position.

\subsubsection{Convergence Analysis of PG-BFGS Iteration}
\label{sec:convergence}

We analyze the convergence of the PG-BFGS iteration using arguments similar 
to those in \cite{XuYin2013}.  Although NPG-BFGS converges faster than 
PG-BFGS empirically, it is not easy to analyze its convergence due to 
\gls{NPG}'s Nesterov's acceleration step and adaptive step size.  In this 
section, we denote the sequence of PG-BFGS iterates by 
$\bigl\{\bigl(\balpha^{(i)},\bcI^{(i)}\bigr)\bigr\}_{i=0}^\infty$.

The following lemma establishes the monotonicity of the PG-BFGS iteration 
for step sizes $\step{\beta}{i}$ that satisfy the majorization condition, 
which includes the above step-size selection as well.

\begin{lem}
  \label{lemma:monotone}
  For a sequence  
  $\bigl\{\bigl(\balpha^{(i)},\bcI^{(i)}\bigr)\bigr\}_{i=0}^\infty$ of 
  PG-BFGS iterates with step size $\beta^{(i)}$ satisfying the majorization 
  condition \eqref{eq:stepCond}, the objective function $f(\balpha,\bcI)$ in 
  \eqref{eq:obj} is monotonically non-increasing:
  \begin{equation}
    f\bigl(\balpha^{(i)},\bcI^{(i)}\bigr) \leq
    f\bigl(\balpha^{(i-1)},\bcI^{(i-1)}\bigr)
    \label{eq:lem23}
  \end{equation}
  for all $i$ under the conditions specified by Theorem~\ref{th:biConvex}.
\end{lem}
\begin{IEEEproof}
  Due to Theorem~\ref{th:biConvex}, $f(\balpha,\bcI^{(i-1)})$ is a convex 
  function of $\balpha$; hence, we can apply 
  \cite[Lemma~2.3]{Beck2009FISTA} to obtain
  \begin{subequations}
    \begin{equation}
      f\bigl(\balpha,\bcI^{(i-1)}\bigr) - 
      f\bigl(\balpha^{(i)},\bcI^{(i-1)}\bigr)
      \geq
      \frac{1}{2\beta^{(i)}}
      \SBR{
        \normbig{\balpha^{(i)}-\wbalpha^{(i)}}_2^2
        +2(\wbalpha^{(i)}-\balpha)^T(\balpha^{(i)}-\wbalpha^{(i)})
      }
      \label{eq:lem23proof}
    \end{equation}
    for any $\balpha$, as long as \eqref{eq:stepCond} is satisfied.  In the 
    PG-BFGS iteration, the momentum term in \eqref{eq:nesterov} is zero, 
    i.e., \eqref{eq:nomomentum} holds.  Plug in \eqref{eq:nomomentum} and
    $\balpha=\balpha^{(i-1)}$ into \eqref{eq:lem23proof}:
    \begin{IEEEeqnarray}{rCl}
      \label{eq:gap}
      f\bigl(\balpha^{(i-1)},\bcI^{(i-1)}\bigr) - 
      f\bigl(\balpha^{(i)},\bcI^{(i-1)}\bigr)
      \geq
      \frac{1}{2\beta^{(i)}}
      \normbig{\balpha^{(i)}-\balpha^{(i-1)}}_2^2
      \geq0
    \end{IEEEeqnarray}
  \end{subequations}
  and \eqref{eq:lem23} follows by using
  the fact $f\bigl(\balpha^{(i)},\bcI^{(i-1)}\bigr)\geq 
  f\bigl(\balpha^{(i)},\bcI^{(i)}\bigr)$, see \ref{bcIStep}.
\end{IEEEproof}

Unfortunately, the monotonicity does not carry over when Nesterov 
acceleration is employed and the momentum term is not zero; this case 
requires further adjustments such as restart, as described in 
\ref{alphaStep}.

Since our $f(\balpha,\bcI)$ are lower bounded [which is easy to argue, see 
Appendix~\ref{app:KL_PropertyProofConv}], the sequence 
$f(\balpha^{(i)},\bcI^{(i)})$ converges.  It is also easy to conclude that 
the sequence 
$a_i\df\frac{1}{\beta^{(i)}}\normbig{\balpha^{(i)}-\balpha^{(i-1)}}_2^2$ is 
Cauchy by showing $\sum_{i=0}^\infty a_i<+\infty$; thus $\balpha^{(i)}$ 
converges to a critical point (also called stationary point) 
$\balpha^\star$ \cite[Lemma~2.2]{AttouchBolte2010}.  A better result 
$\sum_{i=0}^\infty\normbig{\balpha^{(i)}-\balpha^{(i+1)}}_2< +\infty$ 
\cite{XuYin2013} can be established if we can show that $f(\balpha,\bcI)$ 
satisfies the \gls{KUL} property \cite{AttouchBolte2010}, which regularizes 
the (sub)gradient of a function through its value at certain point or the 
whole domain and also ensures the steepness of the function around the 
optimum so that the length of the gradient trajectory is bounded.  The 
\gls{KUL} property has been first used in \cite{AttouchBolte2010} to 
establish the critical-point convergence for an alternating 
proximal-minimization method, which  is then extended by \cite{XuYin2013} 
to the more general block coordinate-descent method.

Now, we can make the following claim on the convergence of PG-BFGS 
iteration.
\begin{thm}
  \label{th:conv}
  Consider the sequence  
  $\bigl\{\bigl(\balpha^{(i)},\bcI^{(i)}\bigr)\bigr\}_{i=0}^\infty$ of 
  PG-BFGS iterates, with step size $\beta^{(i)}$ satisfying the majorization 
  condition \eqref{eq:stepCond}.
  Assume that
  \begin{enumerate}
    \item there exist positive $\beta_+>\beta_->0$ such that 
      $\beta^{(i)}\in[\beta_-,\beta_+]$ for all $i$,
    \item $\cL(\balpha,\bcI)$ is a strong convex function of $\bcI$, and
    \item the first derivative of $f(\balpha,\bcI)$ over $(\balpha,\bcI)$ is 
      Lipschitz continuous
  \end{enumerate}
  and that the conditions of Theorem~\ref{th:biConvex} hold, ensuring the 
  biconvexity of $f$.   Then,
  $(\balpha^{(i)},\bcI^{(i)})$ converges to one of the critical points
  $(\balpha^\star,\bcI^\star)$ of $f(\balpha,\bcI)$ and
  \begin{IEEEeqnarray}{C"C}
    \sum_{i=1}^{\infty}
    \bigl\|\balpha^{(i+1)}-\balpha^{(i)}\bigr\|_2 < +\infty, &
    \sum_{i=1}^{\infty}
    \bigl\|\bcI^{(i+1)}-\bcI^{(i)}\bigr\|_2< +\infty
  \end{IEEEeqnarray}
\end{thm}
\begin{IEEEproof}
  See Appendix~\ref{app:KL_PropertyProofConv}.
\end{IEEEproof}

The conditions for strong convexity of $\cL\PARENS{\balpha,\bcI}$ as a 
function of $\balpha$ are discussed in Sections~\ref{sec:NLLI} and 
\ref{sec:NLLIPois}, see also Section~\ref{sec:rankofA}.  The \gls{KUL} 
property can provide guarantees on the convergence rate under additional 
assumptions, see \cite[Theorem~3.4]{AttouchBolte2010}.
The convergence properties of NPG-BFGS are of great interest because 
NPG-BFGS converges faster than PG-BFGS; establishing these properties is 
left as future work.

\section{Numerical Examples}
\label{sec:numex}

We now evaluate our proposed algorithm by both numerical simulations and 
real data reconstructions.  Here, we focus on the Poisson measurement model 
only and refer to \cite{gdqnde14} for the lognormal case.

We construct the fan-beam X-ray projection transform matrix $\Phi$ and its 
adjoint operator $\Phi^T$ directly on \gls{GPU} with full circular mask 
\cite{dgq11}, and the multi-thread version on CPU is also available; see 
\url{https://github.com/isucsp/imgRecSrc}.

\begin{subequations}
  Before applying the reconstruction algorithms, the measurements $\bcE$ are 
  normalized by division of their maximum such that $\max_n\cE_n=1$, which 
  stabilizes the magnitude of \gls{NLL} by scaling it with a constant, see 
  \eqref{eq:poissNLL}.
  We set the B1-spline tuning constants \eqref{eq:B1tuning} to satisfy
  [see also \eqref{eq:qJcond}]
  \begin{IEEEeqnarray}{c"c"c}
    q^J=10^3, & \kappa_{\lceil 0.5 (J+1) \rceil}=1, &  J=30
  \end{IEEEeqnarray}
  which ensure sufficient coverage (three orders of magnitude) and 
  resolution (30 basis functions) of the basis-function representation of 
  the mass-attenuation spectrum and centering its support around 1.
  We set the convergence and adaptive step-size tuning constants for the 
  NPG-BFGS method as
  \begin{IEEEeqnarray}{c"c}
    (\epsilon, \eta_\balpha, \eta_\bcI, n_\text{sub}) =
    \bigl(10^{-6}, 10^{-3}, 10^{-2}, 20 \bigr), &
    (n,\xi)=(4,0.5)
  \end{IEEEeqnarray}
  limit the number of outer iterations to $4000$ at most.
\end{subequations}

\subsection{Reconstruction from simulated Poisson measurements}
\label{sec:simulatedXrayex}

\begin{figure*}
  \def\width{0.36}
  \centering
  {\begin{subfigure}[t]{0.26\textwidth}
    \centering
    {\includegraphics[width=\textwidth]{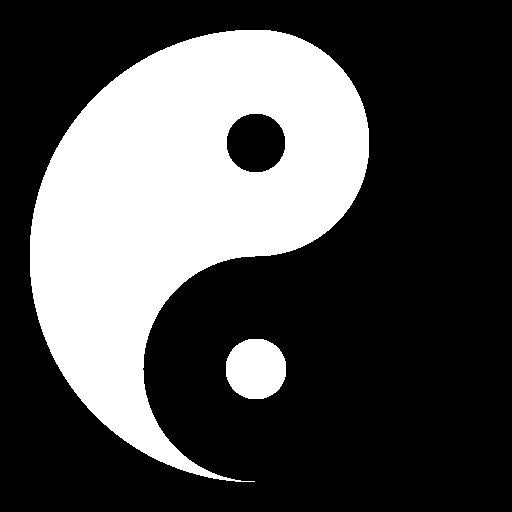}}
    \caption{}
    \label{fig:yangTrue}
  \end{subfigure}}
  \begin{subfigure}[t]{\width\textwidth}
    \centering
    {\includegraphics{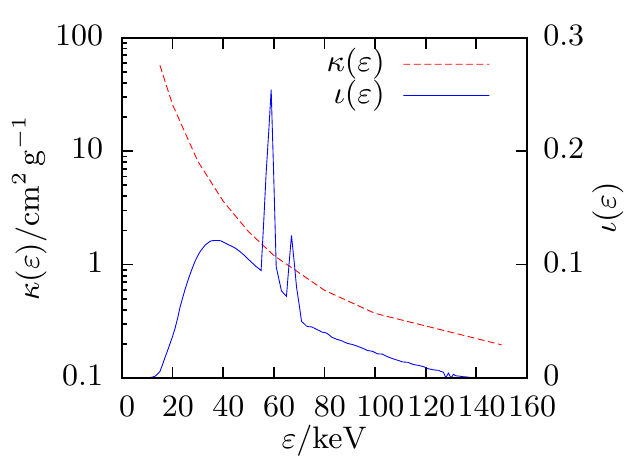}}
    \caption{}
    \label{fig:macAndIota}
  \end{subfigure} \caption{(a) Density-map image and (b) mass attenuation 
    and incident X-ray spectrum as functions of the
    photon energy $\varepsilon$.
  }
\end{figure*}

Consider reconstruction of the \num{512 x 512} image in 
Fig.~\ref{fig:yangTrue} of an iron object with density map 
$\balpha_{\text{true}}$.  We generated a fan-beam polychromatic sinogram, 
with distance from X-ray source to the rotation center equal to \num{2000} 
times the pixel size, using the interpolated mass attenuation 
$\kappa(\varepsilon)$ of iron \cite{MassAttenCoeffNIST} and incident 
spectrum $\iota(\varepsilon)$ from tungsten anode X-ray tubes at 
\SI{140}{\text{\kilo\electronvolt}} with \SI{5}{\percent} relative voltage 
ripple \cite{Boone1997}, see Fig.~\ref{fig:macAndIota}. The 
mass-attenuation spectrum
$\upiota(\kappa)$ is constructed by combining $\kappa(\varepsilon)$ and 
$\iota(\varepsilon)$, see \cite{GuDogandzic2013} and  \cite[Fig.\ 
\ref{qnde14-fig:triRelate}]{gdqnde14}.  Our simulated approximation of the 
noiseless measurements uses \num{130} equi-spaced discretization points 
over the range \SIrange{20}{140}{\text{\kilo\electronvolt}}.
We simulated independent Poisson measurements $\PARENS{\cE_n}_{n=1}^N$ with 
means $\PARENS{\Exp\cE_n}_{n=1}^N=\bcI^\tout(\balpha,\bcI)$.
We mimic real X-ray \gls{CT} system calibration
by scaling the projection matrix $\Phi$ and the spectrum 
$\iota(\varepsilon)$ so that the maximum and minimum of the noiseless 
measurements $\PARENS{\Exp\cE_n}_{n=1}^N$ are $2^{16}$ and $20$, 
respectively. Here, the scales of $\Phi$ and $\iota(\varepsilon)$ 
correspond to the real size that each image pixel represents and the 
current of the electrons hitting the tungsten anode as well as the overall 
scanning time.

Our goal here is to reconstruct a density map of size \num{512 x 512} by 
the measurements from an energy-integrating detector array of size 512 for 
each projection.

Since the true density map is known, we adopt \gls{RSE} as the main metric 
to assess the performance of the compared algorithms:
\begin{equation}
  \RSE\{\what{\balpha}\} =
  1- \PARENS{
    \frac{\what{\balpha}^T \balpha_{\text{true}}}
    {{\|\what{\balpha}\|_2 \|\balpha_{\text{true}}\|_2}}
  }^2
  \label{eq:defRSE}
\end{equation}
where $\balpha_\text{true}$ and $\what{\balpha}$ are the true and 
reconstructed signals, respectively. Note that \eqref{eq:defRSE} is 
invariant to scaling $\what{\balpha}$ by a nonzero constant, which is 
needed because the magnitude level of $\balpha$ is not identifiable due to 
the ambiguity of the density map and mass attenuation spectrum (see also 
Section.~\ref{sec:ambiguity}).

We compare
\begin{itemize}
  \item the traditional \gls{FBP} method without \cite[Ch.~3]{Kak1988CT} 
    and with linearization \cite{Herman1979}, i.e., based on the `data' 
    \begin{subequations}
      \begin{IEEEeqnarray}{c"c}
        \label{eq:wolin}
        \by = -{\ln_\circ(\bcE)} & \text{(without linearization)}
 \\
        \label{eq:wlin}
        \by=\bigl(\upiota^\tL\bigr)_\circ^{-1}(\bcE) & \text{(with 
        linearization)}
      \end{IEEEeqnarray}
    \end{subequations}
    respectively,
  \item linearized \gls{BPDN} in which we apply the NPG approach to solve 
    the \gls{BPDN} problem \cite{Beck2009TV}:
    \begin{IEEEeqnarray}{c}
      \label{eq:linNPGobjectivefunction}
      \min_{\balpha} \frac{1}{2} \|\by-\Phi \balpha\|_2^2 + u' r(\balpha)
    \end{IEEEeqnarray}
    where $\by$ are the linearized measurements by \eqref{eq:wlin} and the 
    penalty $r(\balpha)$ has been defined in \eqref{eq:r}.
  \item our
    \begin{itemize}
      \item NPG-BFGS alternating descent method,
      \item \gls{NPG} for known mass attenuation spectrum 
        $\upiota(\kappa)$.
    \end{itemize}
    with the Matlab implementation available at 
    \url{https://github.com/isucsp/imgRecSrc}.
\end{itemize}

For all methods that use sparsity and nonnegativity regularization 
(NPG-BFGS, NPG, and linearized BPDN) the regularization constants $u$ and 
$u'$ have been tuned manually for best \gls{RSE} performance.  All 
iterative algorithms employ the convergence criterion 
\eqref{eq:convConNPGAS} with threshold $\epsilon=\num{e-6}$ and the maximum 
number of iterations set to \num{4000}.  We initialize iterative 
reconstruction schemes with or without linearization using the 
corresponding \gls{FBP} reconstructions.

Here, the \emph{non-blind} linearized FBP, NPG, and linearized BPDN methods  
assume known $\upiota(\kappa)$ [i.e., known incident spectrum of the X-ray 
machine and mass attenuation (material)], which we computed using 
\eqref{eq:upiota} with $\bcI$ equal to the exact sampled $\upiota(\kappa)$ 
and $J=100$ spline basis functions spanning three orders of magnitude.
The \gls{FBP} and  NPG-BFGS methods are \emph{blind} and do not assume 
knowledge of $\upiota(\kappa)$;  \gls{FBP} ignores the polychromatic source 
effects whereas NPG-BFGS corrects blindly for these effects.

\begin{figure}
  \centering
  \includegraphics{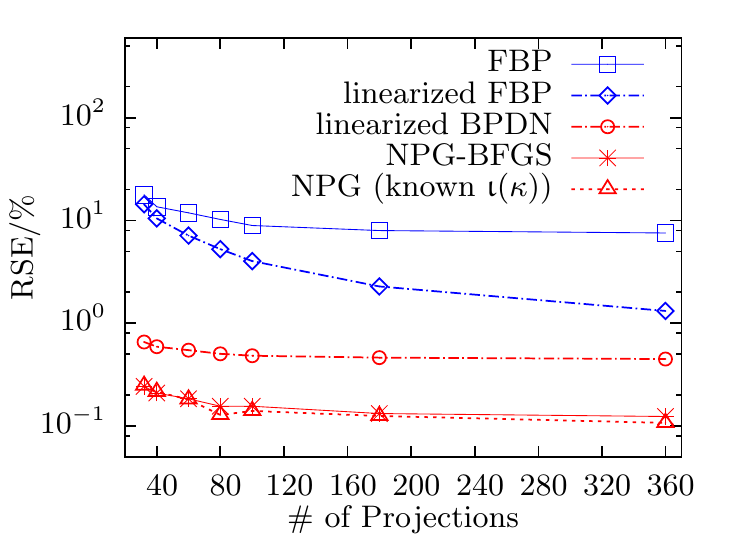}
  \caption{
    Average \glspl{RSE} as functions of the number of projections.}
  \label{fig:rmse_prj_yang}
\end{figure}

Fig.~\ref{fig:rmse_prj_yang} shows the average \glspl{RSE} (over 5 Poisson 
noise realizations) of different methods as functions of the number of 
fan-beam projections in the range from \SIrange{0}{359}{\degree}.  
\glspl{RSE} of the methods that do not assume knowledge of the mass 
attenuation spectrum $\upiota(\kappa)$ are shown using solid lines whereas 
dashed lines represent methods that assume known $\upiota(\kappa)$ [i.e., 
known incident spectrum of the X-ray machine and mass attenuation 
(material)].
Blue color presents methods that do not employ regularization; red color 
presents methods that employ regularization.

FBP ignores the polychromatic nature of the measurements and consequently 
performs poorly and does not improve as the number of projections 
increases.  Linearized \gls{FBP}, which assumes perfect knowledge of the 
mass attenuation spectrum, performs much better than FBP as shown in 
Fig.~\ref{fig:rmse_prj_yang}.  Thanks to the nonnegativity and sparsity 
that it imposes, linearized \gls{BPDN} achieves up to 20 times smaller 
\gls{RSE} than linearized \gls{FBP}.  However, the linearization process 
employed by linearized \gls{BPDN} enhances the noise (due to the 
zero-forcing nature of linearization) and does not account for the Poisson 
measurements model.  Note that linearized \gls{BPDN} exhibits a noise floor 
as the number of projections increases.

As expected, \gls{NPG} is slightly better than \gls{NPG}-BFGS because it 
assumes knowledge of $\upiota(\kappa)$.  NPG and NPG-BFGS attain  
\glspl{RSE} that are \SIrange{24}{37}{\percent} that of linearized 
\gls{BPDN}, which can be attributed to optimal statistical processing of 
the proposed methods, in contrast with suboptimal linearization.  
\glspl{RSE} of \gls{NPG} and NPG-BFGS reach a noise floor when the number 
of projections increases beyond 180.

\begin{figure*}
  \def\width{0.32}
  \def\height{140}
  \def\slen{2}
  \centering
  \begin{subfigure}[t]{\width\textwidth}
    \centering
    \includegraphics[width=\textwidth]{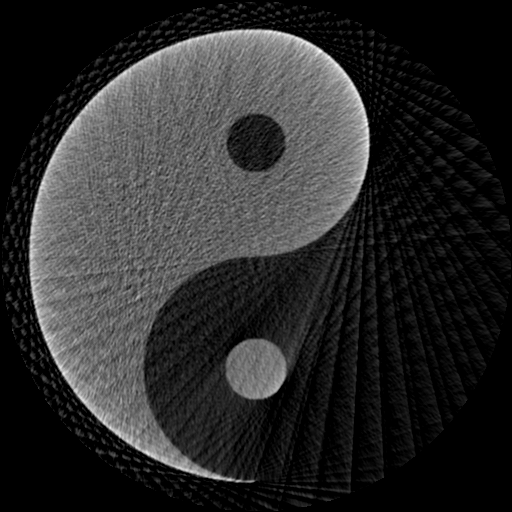}
    \put(-78,0){\color{red}\rule{0.5pt}{150pt}}
    \putNumberT{11.83}{\height}{\slen}
    \caption{FBP}
    \label{fig:fbp_yang}
  \end{subfigure}
  \begin{subfigure}[t]{\width\textwidth}
    \centering
    \includegraphics[width=\textwidth]{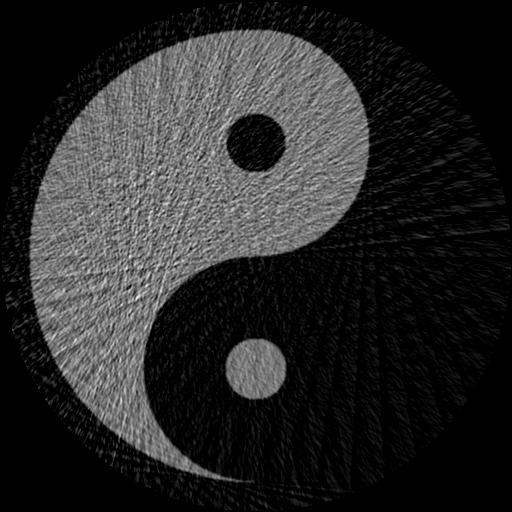}
    \putNumberT{7.12}{\height}{\slen}
    \caption{linearized FBP}
    \label{fig:linFBP_yang}
  \end{subfigure}
  \begin{subfigure}[t]{\width\textwidth}
    \centering
    \includegraphics[width=\textwidth]{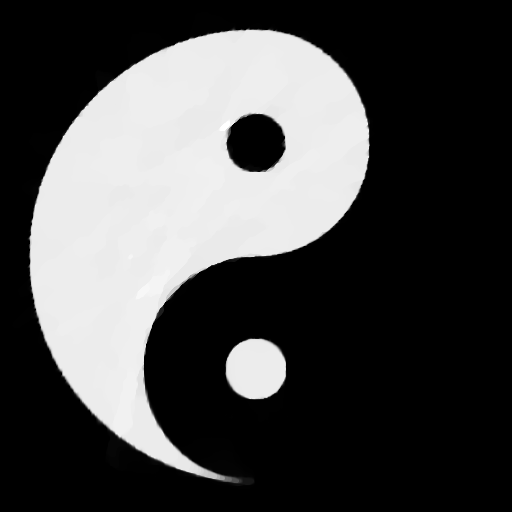}
    \putNumberT{0.55}{\height}{\slen}
    \caption{linearized BPDN}
    \label{fig:linNPGTV_yang}
  \end{subfigure}
  \\
   \begin{subfigure}[t]{\width\textwidth}
    \centering
    \includegraphics[width=\textwidth]{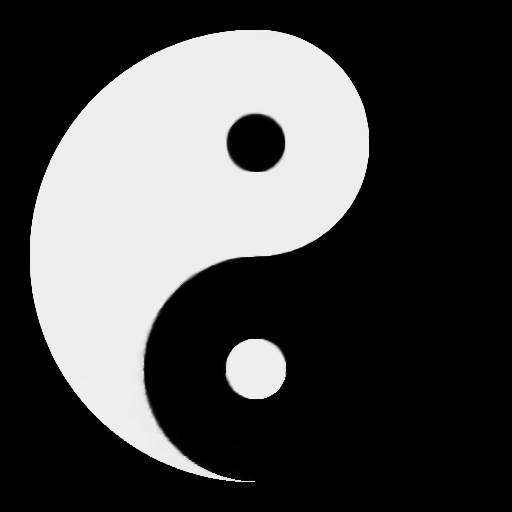}
    \putNumberT{0.18}{\height}{\slen}
    \caption{NPG-BFGS}
    \label{fig:npgTV_yang}
  \end{subfigure}
  \begin{subfigure}[t]{\width\textwidth}
    \centering
    \includegraphics[width=\textwidth]{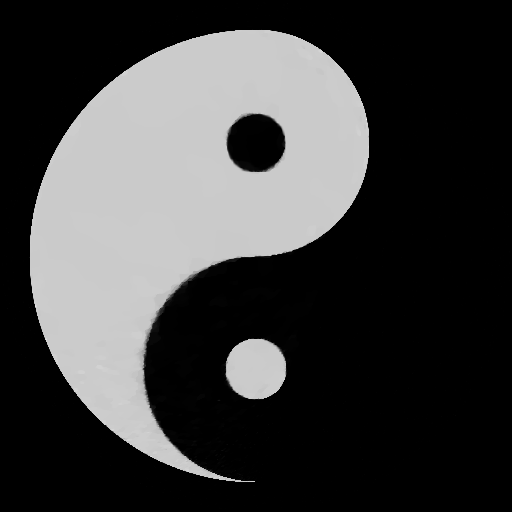}
    \putNumberT{0.18}{\height}{\slen}
    \caption{NPG (known $\upiota(\kappa)$)}
    \label{fig:npgTValpha_yang}
  \end{subfigure}
   \caption{Reconstructions from \num{60} projections.}
  \label{fig:yang}
\end{figure*}

Fig.~\ref{fig:yang} shows the reconstructions from 60 equi-spaced fan-beam 
projections with spacing $6^\circ$.\footnote{Since the density-map 
  reconstruction can only have nonnegative elements, we show the 
  reconstructions in a color-map ranging from 0 to its maximum estimated 
  pixel value, which effectively does the negative truncation for the 
  (linearized) \gls{FBP} methods where the nonnegativity signal constraints 
are not enforced.}
The \gls{FBP} reconstruction in Fig.~\ref{fig:fbp_yang} is corrupted by 
both aliasing and beam-hardening (cupping and streaking) artifacts.  The 
intensity is high along the outside-towards boundary, the ``eye'' and the 
non-convex ``abdomen'' area of the ``fish'' has positive components, and 
the ball above the ``tail'' has a streak artifact connecting to the 
``fish'' head.  Linearized \gls{FBP} removes the beam-hardening artifacts, 
but retains the aliasing artifacts and enhances the noise due to the 
zero-forcing nature of linearization, see Fig.~\ref{fig:linFBP_yang}.  Upon 
enforcing the nonnegativity and sparsity constraints, linearized \gls{BPDN} 
removes the negative components and achieves a smooth reconstruction with a 
\SI{0.55}{\percent} \gls{RSE}.  Thanks to the superiority of the proposed 
model that accounts for both the polychromatic X-ray and Poisson noise in 
the measurements, \gls{NPG}-BFGS and \gls{NPG} achieve the best 
reconstructions, see Fig.~\ref{fig:npgTValpha_yang} and 
\ref{fig:npgTV_yang}.

\begin{figure*}
  \def\width{0.46}
  \centering
  {\begin{subfigure}[t]{0.46\textwidth}
    \centering
    \includegraphics{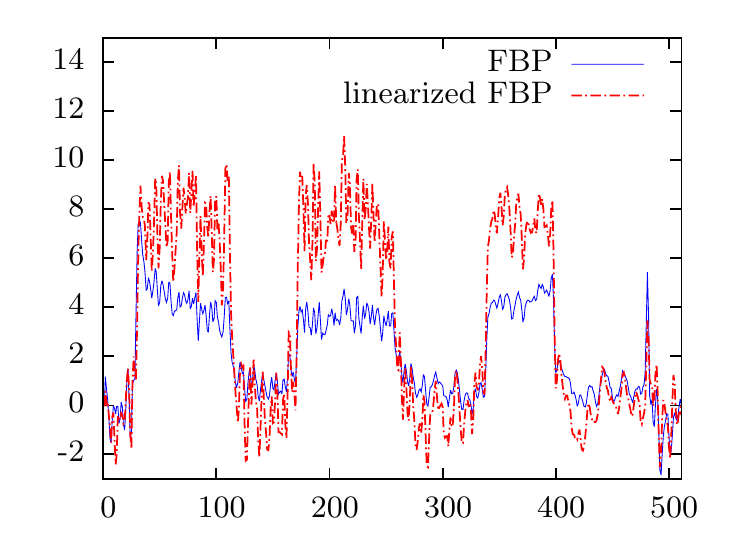}
    \caption{}
    \label{fig:profile_yang}
  \end{subfigure}}
  {\begin{subfigure}[t]{0.46\textwidth}
    \centering
    \includegraphics{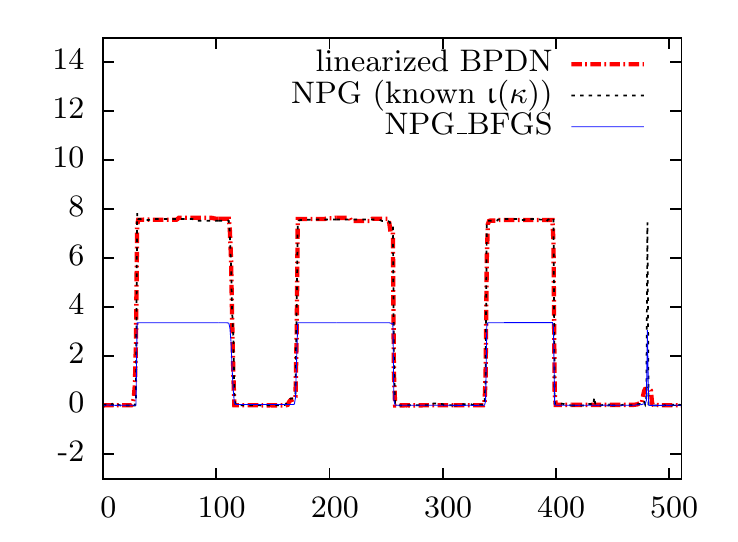}
    \caption{}
    \label{fig:profile_yang1}
  \end{subfigure}}
  {\begin{subfigure}[t]{0.46\textwidth}
    \centering
    \includegraphics{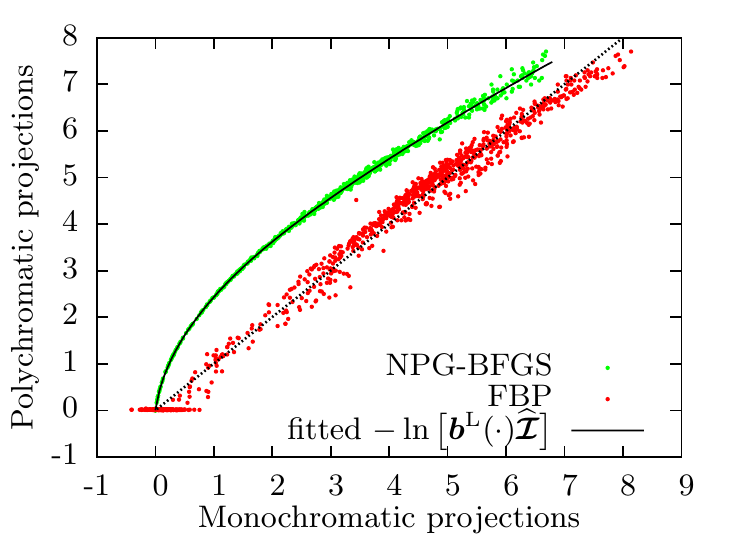}
    \caption{}
    \label{fig:linearization_yang}
  \end{subfigure}}
  \label{eq:yangPlot}
  \caption{(a)--(b) The 250th-column profiles of the reconstructions, and 
  (c) the polychromatic measurements as function of the monochromatic 
projections and corresponding fitted curves.}
\end{figure*}

Figs.~\ref{fig:profile_yang} and \ref{fig:profile_yang1} show the profiles 
of each reconstruction at the 250th column indicated by the red line in 
Fig.~\ref{fig:fbp_yang}.
In Fig.~\ref{fig:linearization_yang}, we show the scatter plots with 1000 
randomly selected points representing \gls{FBP} and \gls{NPG}-BFGS 
reconstructions of the \texttt{C-II} object from 60 fan-beam projections.
 Denote by $\bigl(\what{\balpha},\what{\bcI}\bigr)$ the estimate of 
 $(\balpha,\bcI)$ obtained upon convergence of the NPG-BFGS iteration.
The $y$-coordinates in the scatter plots in 
Fig.~\ref{fig:linearization_yang} are the \emph{noisy}  measurements in log 
scale $-{\ln\cE_n}$ and the corresponding $x$-coordinates are the 
monochromatic projections $\bphi_n^T\what{\balpha}_\text{FBP}$ (red) and 
$\bphi_n^T\what{\balpha}$ (green) of the estimated density maps;  
$-{\ln\bigl[\bb^\tL(\cdot)\what{\bcI}\bigr]}$ is the inverse linearization 
function that maps monochromatic projections to fitted \emph{noiseless} 
polychromatic projections $-{\ln\cI_n^\tout}(\what{\balpha},\what{\bcI})$.  
The vertical-direction differences between the NPG-BFGS scatter plot and 
the corresponding linearization curve show goodness of fit between the 
measurements and our model.  Since \gls{FBP} assumes linear relation 
between $-{\ln_\circ\bcI^\tout}$ and $\Phi\balpha$, its scatter plot
(red) can be fitted by a straight line $y=x$, as shown in 
Fig.~\ref{fig:linearization_yang}.
A few points in the \gls{FBP} scatter plot with $\ln\cE_n=0$ and positive 
monochromatic projections indicate severe streaking artifacts.
Observe relatively large residuals with bias, which remain even if more 
sophisticated linear models, e.g., iterative algorithms with sparsity and 
nonnegativity constraints, were adopted, thereby necessitating the need for 
accouting for the polychromatic source; see the numerical examples in 
\cite{gdqnde14}.

Note that the nonnegativity constraints on $\balpha$ are especially 
effective in the process of getting good estimation of $\bcI$.  Without 
these constraints, it is even impossible for \ref{alphaStep} and 
\ref{bcIStep} to converge to reasonable place.

\subsection{X-ray \gls{CT} Reconstruction from Real Data}
\label{sec:realXrayCTex}

\begin{figure*}
  \def\width{0.42}
  \centering
  \begin{subfigure}[t]{\width\textwidth}
    \centering
    \includegraphics[width=\textwidth]{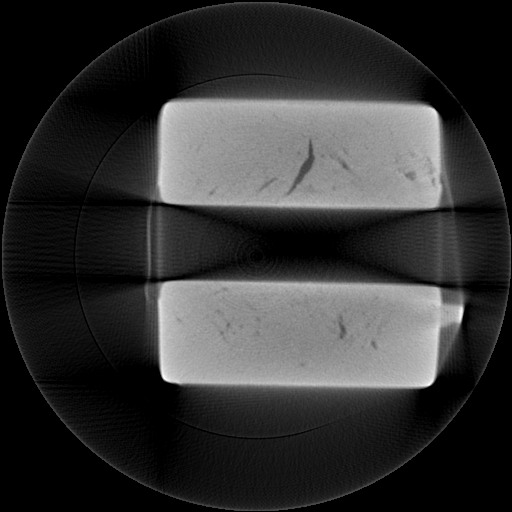}
    \caption{FBP}
    \label{fig:fbp_cpm}
  \end{subfigure}
  \begin{subfigure}[t]{\width\textwidth}
    \centering
    \includegraphics[width=\textwidth]{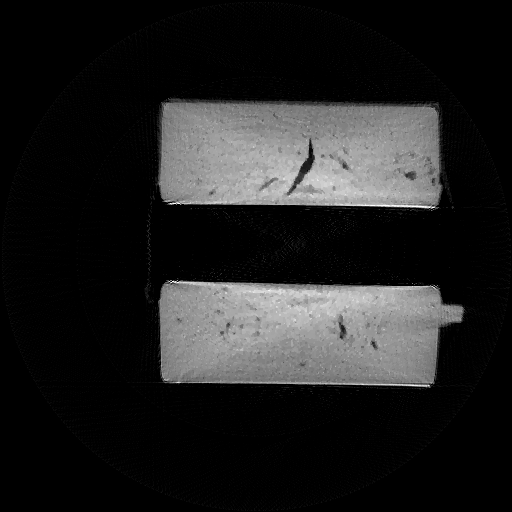}
    \caption{NPG-BFGS}
    \label{fig:npgTV_cpm}
  \end{subfigure}
  \begin{subfigure}[t]{\width\textwidth}
    \centering
    \includegraphics[width=\textwidth]{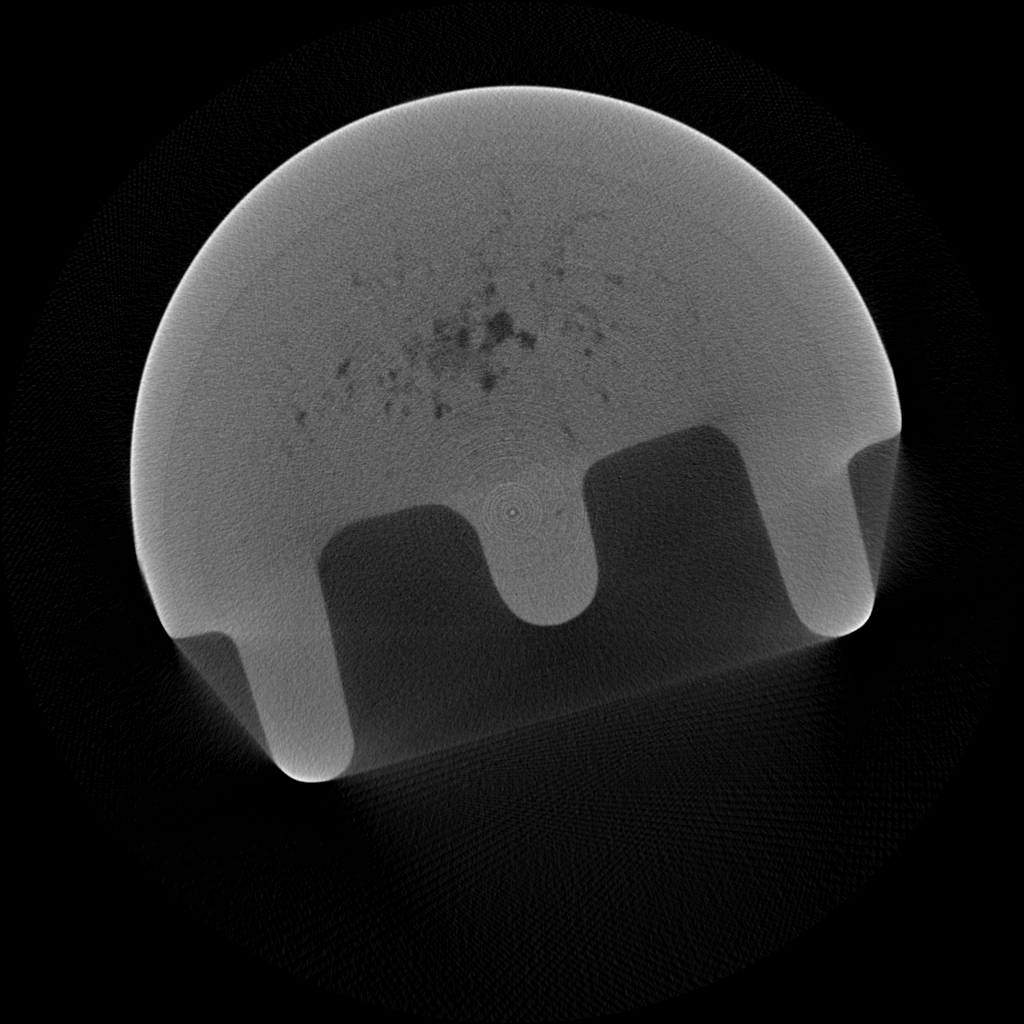}
    \put(-197,132.17){\color{red}\rule{197pt}{0.2pt}}
    \put(-197,95.23){\color{red}\rule{197pt}{0.2pt}}
    \caption{FBP}
    \label{fig:fbp_casting}
  \end{subfigure}
  \begin{subfigure}[t]{\width\textwidth}
    \centering
    \includegraphics[width=\textwidth]{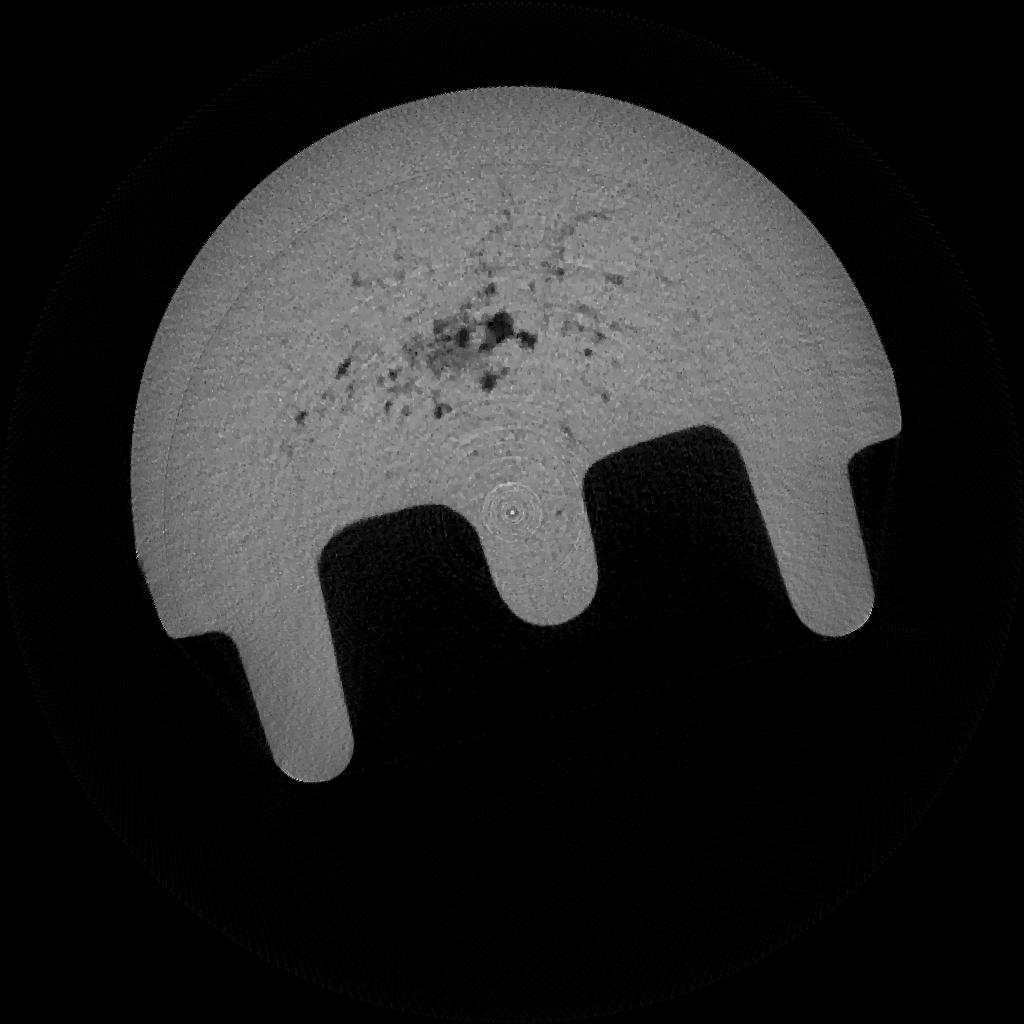}
    \put(-197,132.17){\color{red}\rule{197pt}{0.2pt}}
    \put(-197,95.23){\color{red}\rule{197pt}{0.2pt}}
    \caption{NPG-BFGS}
    \label{fig:npgTV_casting}
  \end{subfigure}
  \begin{subfigure}[b]{\width\textwidth}
    \centering
    \includegraphics[width=\textwidth]{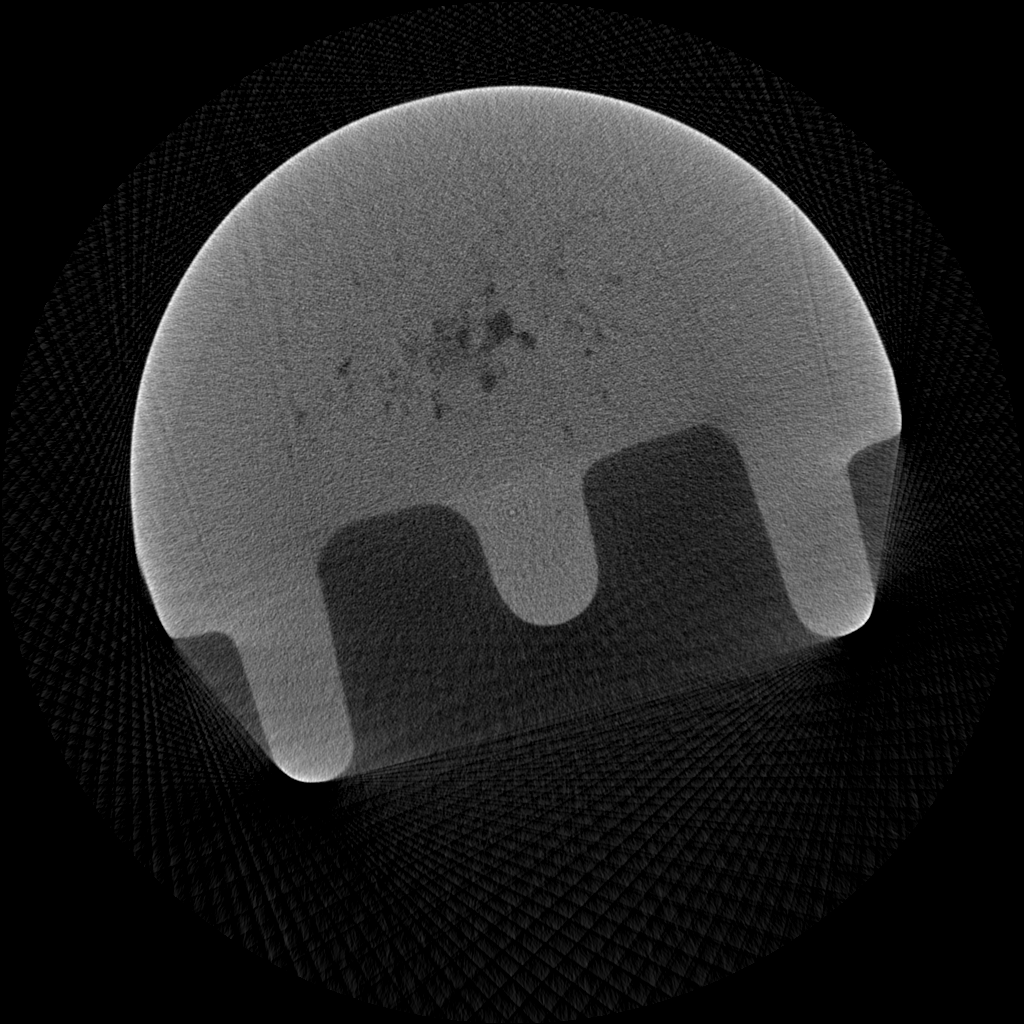}
    \caption{FBP (120 projections)}
    \label{fig:fbp_casting_120}
  \end{subfigure}
  \begin{subfigure}[b]{\width\textwidth}
    \centering
    \includegraphics[width=\textwidth]{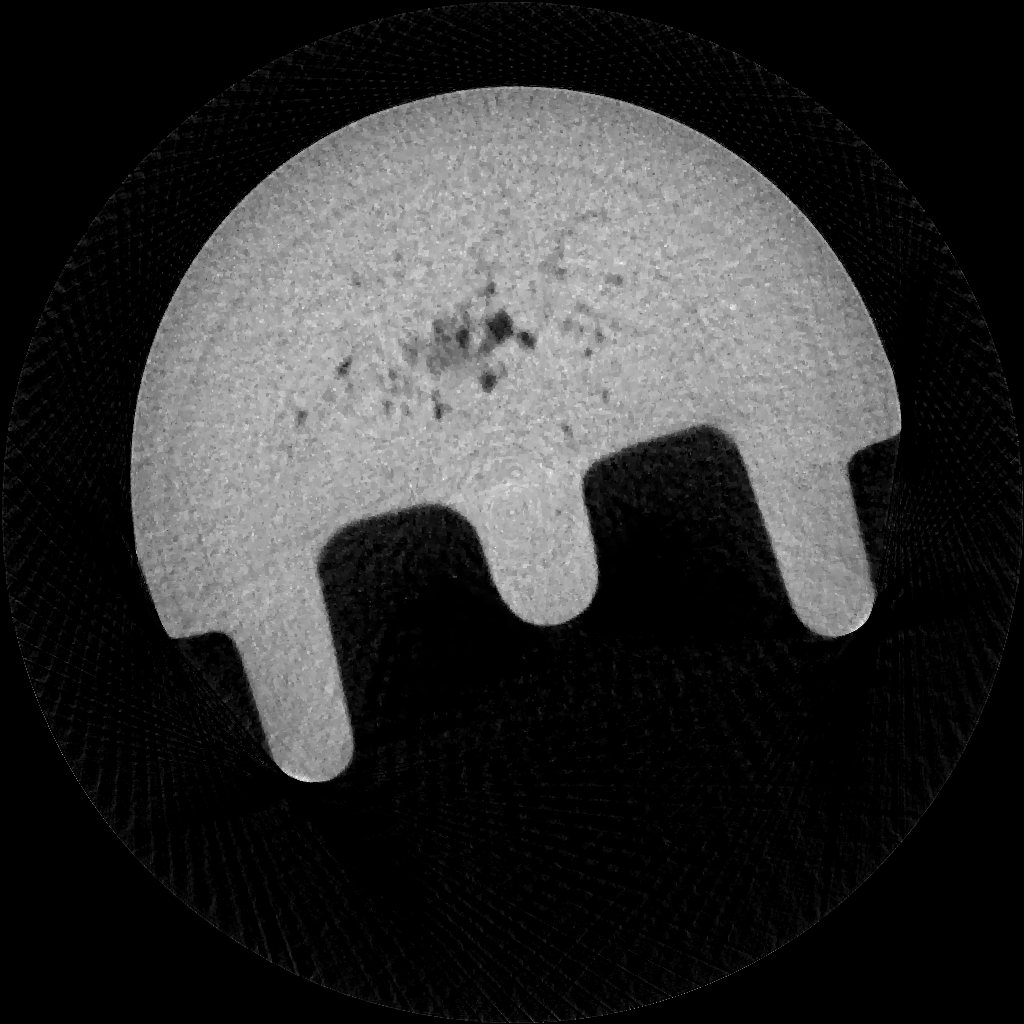}
    \caption{NPG-BFGS (120 projections)}
    \label{fig:npgTV_casting_120}
  \end{subfigure}
  \caption{ \texttt{C-I} and \texttt{C-II}  object reconstructions from 
  fan-beam projections using the \gls{FBP} and NPG-BFGS methods.}
  \label{fig:realdata}
\end{figure*}

We compare the NPG-BFGS and linear \gls{FBP} methods by applying them to 
reconstruct two industrial objects containing defects, labeled \texttt{C-I}  
and \texttt{C-II}, from real fan-beam projections.
Here, NPG-BFGS achieves visually good reconstructions for $u=10^{-5}$, 
presented in Fig.~\ref{fig:realdata}.

The \texttt{C-I} data set consists of \num{360} equi-spaced fan-beam 
projections with \SI{1}{\degree} separation collected using an array of 
\num{694} detectors, with distance of X-ray source to the rotation center 
equal to $3492$ times the detector size. Figs.~\ref{fig:fbp_cpm} and 
\ref{fig:npgTV_cpm} show \num{512 x 512} density-map image reconstructions 
of object \texttt{C-I} using the \gls{FBP} and NPG-BFGS methods, 
respectively.  The linear \gls{FBP} reconstruction, which does not account 
for the polychromatic nature of the X-ray source,  suffers from severe 
streaking (shading that occupies the empty area) and cupping (high 
intensity along the object's border) artifacts whereas the NPG-BFGS 
reconstruction removes these artifacts thanks to accounting for the 
polychromatic X-ray source.

The \texttt{C-II} data set consists of \num{360} equi-spaced fan-beam 
projections with \SI{1}{\degree} separation collected using an array of 
\num{1380} detectors, with distance of X-ray source to the rotation center 
equal to $8696$ times the detector size.  Figs.~\ref{fig:fbp_casting} and 
\ref{fig:npgTV_casting} show \num{1024 x 1024} density-map image 
reconstructions of object \texttt{C-II} by the \gls{FBP} and NPG-BFGS 
methods, respectively.  The NPG-BFGS reconstruction removes the streaking 
and cupping artifacts exhibited by the linear \gls{FBP}, with enhanced 
contrast for both the inner defects and object boundary.  
Figs.~\ref{fig:fbp_casting_120} and \ref{fig:npgTV_casting_120} show the 
\gls{FBP} and NPG-BFGS reconstructions from downsampled \texttt{C-II} data 
set with \num{120} equi-spaced fan-beam projections with \SI{3}{\degree} 
separation.  The \gls{FBP} reconstruction in Fig.~\ref{fig:fbp_casting_120} 
exhibits both beam-hardening and aliasing artifacts. In contrast, the 
NPG-BFGS reconstruction in Fig.~\ref{fig:npgTV_casting_120} does not 
exhibit these artifacts because it accounts for the polychromatic X-ray 
source and employs signal-sparsity regularization in \eqref{eq:r}.  Indeed, 
if we reduce the regularization constant $u$ sufficiently, the aliasing 
effect will occur in the NPG-BFGS reconstruction in 
Fig.~\ref{fig:npgTV_casting_120} as well.

\begin{figure}
  \centering
  {\begin{subfigure}{0.48\textwidth}
    \centering
    \includegraphics{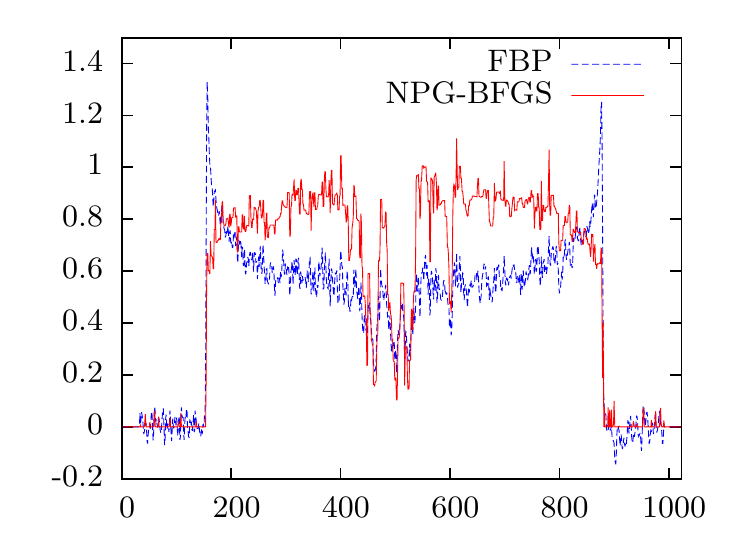}
    \caption{337th row}
    \label{fig:profile_casting}
  \end{subfigure}}
  {\begin{subfigure}{0.48\textwidth}
    \centering
    {\includegraphics{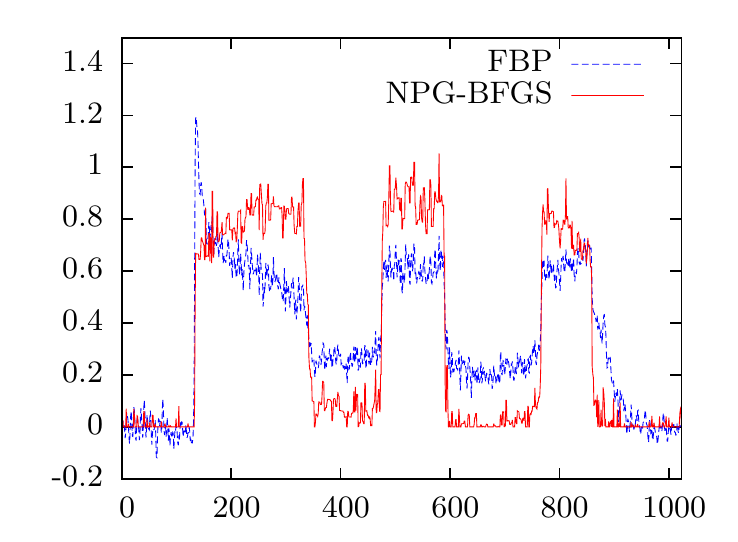}}
    \caption{531st row}
    \label{fig:profile_casting1}
  \end{subfigure}}
  {\begin{subfigure}{0.48\textwidth}
    \centering
    \includegraphics{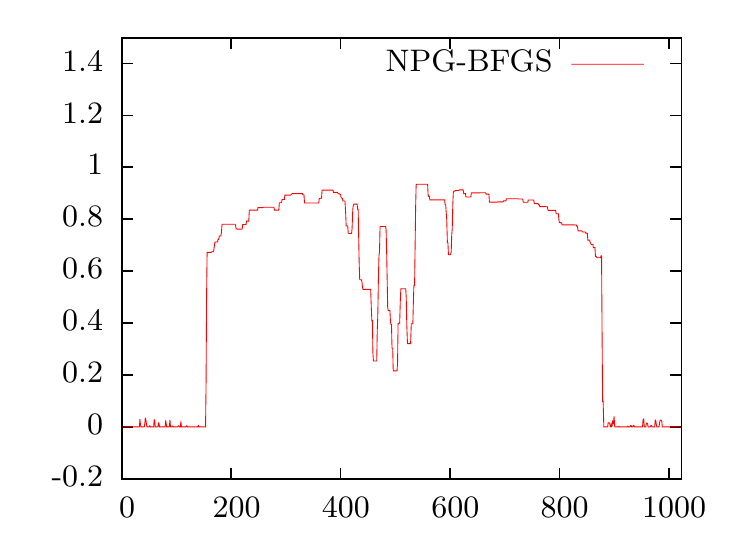}
    \caption{337th row}
    \label{fig:profile_casting2}
  \end{subfigure}}
  {\begin{subfigure}{0.48\textwidth}
    \centering
    \includegraphics{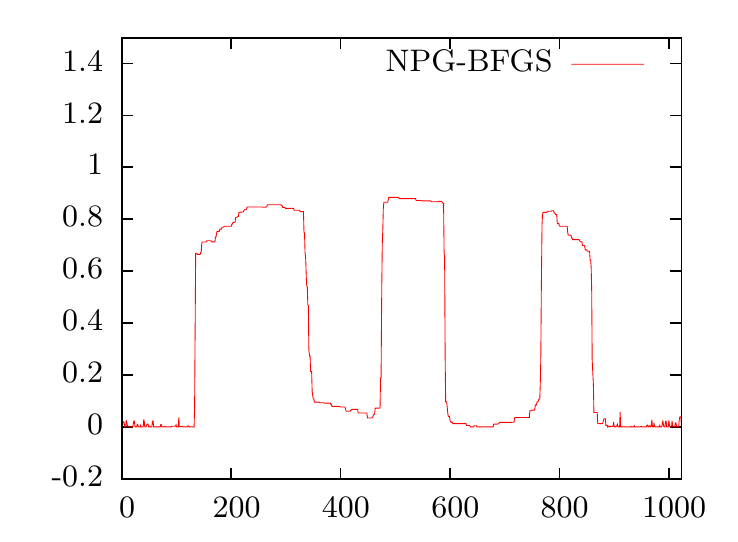}
    \caption{531st row}
    \label{fig:profile_casting3}
  \end{subfigure}}
  \caption{
    \texttt{C-II} object reconstruction profiles with NPG-BFGS 
    reconstructions using the regularization tuning constant (a)--(b) 
    $u=10^{-5}$ and (c)--(d)
    $u=10^{-4}$.}
  \label{fig:profiles}
\end{figure}

Fig.~\ref{fig:profiles} shows the reconstruction profiles of the 337th and 
531th rows highlighted by the red horizontal lines across 
Figs.~\ref{fig:fbp_casting} and \ref{fig:npgTV_casting}.  Noise in the 
NPG-BFGS reconstructions can be reduced by increasing the regularization 
parameter $u$:  Figs.~\ref{fig:profile_casting2} and 
\ref{fig:profile_casting3} show the corresponding NPG-BFGS reconstruction 
profiles for $u=10^{-4}$, which is 10 times that in 
Figs.~\ref{fig:profile_casting} and \ref{fig:profile_casting1}.

Observe that the NPG-BFGS reconstructions of  \texttt{C-I} and  
\texttt{C-II}  have higher contrast around the inner region where cracks 
reside.  Indeed, our reconstructions have slightly higher intensity in the 
center, which is likely due to the detector saturation that lead to 
measurement truncation; other possible causes could be scattering or noise 
model mismatch.  (We have replicated this slight non-uniformity by applying 
our reconstruction to simulated truncated measurements.)  This effect is 
visible in \texttt{C-I} reconstruction in Fig.~\ref{fig:npgTV_cpm} and is 
barely visible in the  \texttt{C-II} reconstruction in 
Fig.~\ref{fig:npgTV_casting}, but can be observed in the profiles in 
Fig.~\ref{fig:profiles}.  We leave further verification of causes and 
potential correction of this problem to future work and note that this 
issue does not occur in simulated-data examples that we constructed, see 
Section~\ref{sec:simulatedXrayex}.

\begin{figure*}
  \def\width{0.42}
  \centering
  \begin{subfigure}[b]{\width\textwidth}
    \centering
    \includegraphics[width=\textwidth]{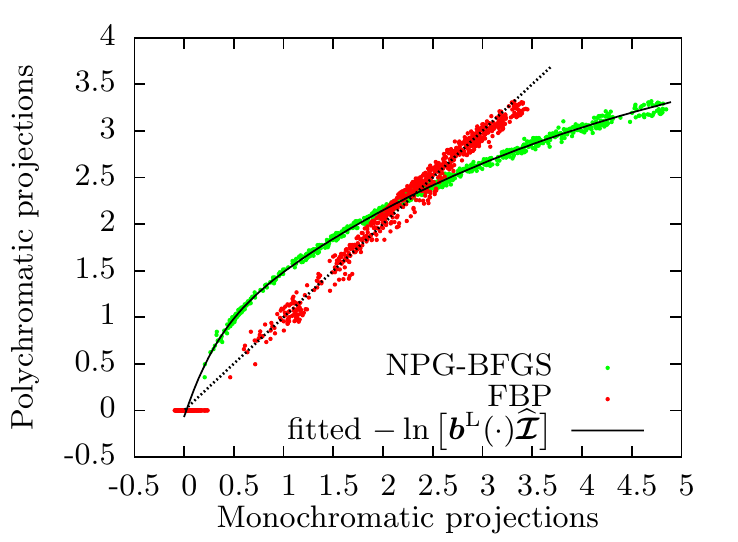}
    \caption{}
    \label{fig:lin}
  \end{subfigure}
  \begin{subfigure}[b]{\width\textwidth}
    \centering
    \includegraphics[width=\textwidth]{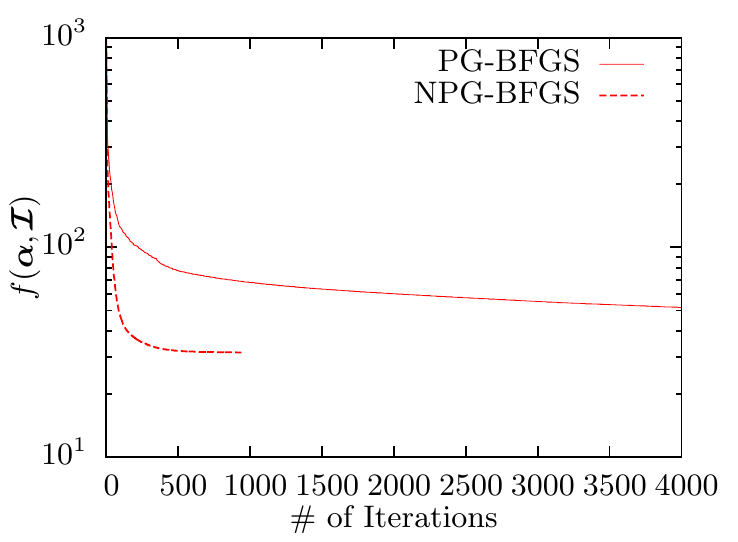}
    \caption{}
    \label{fig:castingTrace}
  \end{subfigure}
  \caption{(a) Polychromatic measurements as functions of monochromatic 
    projections and corresponding fitted inverse linearization curves, and 
    (b) the objective function as a function of the iteration index $i$.
  }
  \label{fig:lin_and_obj}
\end{figure*}

In Fig.~\ref{fig:lin}, we show the scatter plots with 1000 randomly 
selected points representing \gls{FBP} and \gls{NPG}-BFGS reconstructions 
of the \texttt{C-II} object from 360 projections.  A few points in the 
\gls{FBP} scatter plot with $\ln\cE_n=0$ and positive monochromatic 
projections indicate severe streaking artifacts, which we also observed in 
the simulated example in Section~\ref{sec:simulatedXrayex}, see 
Fig.~\ref{fig:linearization_yang}.

We now illustrate the advantage of using Nesterov's acceleration in 
\ref{alphaStep} of our iteration. Fig.~\ref{fig:castingTrace} shows the 
objective $f(\balpha,\bcI)$ as a function of outer iteration index $i$ for 
the NPG-BFGS and PG-BFGS methods
applied to  \texttt{C-II} reconstruction from
360 projections.  Thanks to the Nesterov's acceleration 
\eqref{eq:nesterov}, NPG-BFGS is at least 10 times faster than PG-BFGS, 
which runs until it reaches the maximum-iteration limit.

\section{Conclusion}
\label{sec:conclusion}

We developed a model that requires no more information than the 
conventional \gls{FBP} method but is capable of beam-hardening artifact 
correction.  The proposed model, based on the separability of the 
attenuation, combines the variations of the mass attenuation and X-ray 
spectrum to \emph{mass-attenuation spectrum}.  This model solves the single 
material scenario and convert it to a bi-convex optimization problem.  We 
proposed an alternatively descent iterative algorithm to find the solution.  
We proved the \gls{KUL} properties of the object function under both 
Gaussian and Poisson noise scenarios, and gave primitive conditions that 
guarantees the bi-convexity.  Numerical experiments on both simulated and 
real X-ray CT data are presented.
Our \emph{blind} method for sparse X-ray \gls{CT} reconstruction
\begin{enumerate*}[label=(\roman*)]
  \item does not rely on knowledge of the X-ray source spectrum and the 
    inspected material (i.e., its mass-attenuation function),
  \item \label{blindbetterthanlinearization} matches or outperforms 
    non-blind linearization methods that assume perfect knowledge of the 
    X-ray source and material properties.
\end{enumerate*}
This method is based on our new parsimonious model for polychromatic X-ray 
\gls{CT} that exploits separability of attenuation of the inspected object.  
It is of interest to generalize this model beyond X-ray \gls{CT} so that it 
applies to other physical models and measurement scenarios that have 
separable structure; here, we would construct the most general possible 
formulation of ``separable structure''.  For example,
our probabilistic model in this paper is closely related to \glspl{GLM} in 
statistics \cite{McCullagh1989}, and can be viewed as their regularized 
(nonnegativity- and sparsity-enforcing) generalization as well.  

The underlying optimization problem for performing our blind sparse signal 
reconstruction is \emph{biconvex} with respect to the density map and mass 
density spectrum parameters; solving and analyzing bi- and multiconvex 
problems is of great general interest in optimization theory, see 
\cite{XuYin2013} and references therein. An interesting conjecture is that 
the general version of \ref{blindbetterthanlinearization} holds as a 
consequence of the separability in the measurement model.

Future work will also include extending our parsimonious mass-attenuation 
parameterization to multiple materials and developing
corresponding reconstruction algorithms.

\appendices

\section{Derivation of the Mass-Attenuation Parameterization}
\label{app:anykappa}

\renewcommand{\theequation}{A\arabic{equation}}
\setcounter{equation}{0}

\renewcommand{\thesubsection}{\Alph{section}-\Roman{subsection}}
\renewcommand{\thesubsubsection}{\Alph{section}-\Roman{subsection}
\arabic{subsubsection}}
\renewcommand {\thesubsectiondis}{\Roman{subsection}}

Observe that the mass attenuation $\kappa(\varepsilon)$ and incident
energy density $\iota(\varepsilon)$ are both functions of
$\varepsilon$, see Fig.~\ref{fig:triRelate}.  Thus, to combine the
variation of these two functions and reduce the degree of freedom, we
rewrite $\iota(\varepsilon)$ as a function of $\kappa$ and
set $\kappa$ as the integral variable.  

All mass-attenuation functions $\kappa(\varepsilon)$ encountered in 
practice can be divided into piecewise-continuous segments, where each 
segment is a differentiable monotonically decreasing function of 
$\varepsilon$, see \cite[Tables 3 and 4]{MassAttenCoeffNIST} and 
\cite[Sec.\ 2.3]{Huda2009}.  The points of discontinuity in 
$\kappa(\varepsilon)$ are referred to as $K$-edges and are caused by the 
interaction between photons and $K$ shell electrons, which  occurs only 
when $\varepsilon$ reaches the binding energy of the $K$ shell electron.
One example in Fig.~\ref{fig:ironKappa} is the mass attenuation coefficient 
curve of iron with a single $K$-edge at 
\SI{7.11}{\text{\kilo\electronvolt}}.

If the energy range used in a \gls{CT} scan \emph{does not} contain any of 
the $K$-edges, $\kappa(\varepsilon)$ becomes monotonic and invertible, 
which we consider first in Appendix~\ref{app:invertiblekappa}.
This simple case, depicted in Fig.~\ref{fig:triRelate}, gives the main idea 
behind our simplified model.

In Appendix~\ref{app:piecewisecontinuouskappa},
we then consider the general case where the energy range used in a \gls{CT} 
scan contains $K$-edges.

\subsection{Invertible $\kappa(\varepsilon)$}
\label{app:invertiblekappa}

Consider first the simple case where $\kappa(\varepsilon)$ is monotonically 
decreasing and invertible, as depicted in Fig.~\ref{fig:triRelate},
and define the differentiable inverse of $\kappa(\varepsilon)$ as 
$\varepsilon(\kappa)$.  The change of variables 
$\varepsilon=\varepsilon(\kappa)$ in the integral
expressions \eqref{eq:constanttotalenergy} and 
\eqref{eq:polyModelseparable}
yields
\begin{subequations}
  \label{eq:massattenuationparametrization}
  \begin{IEEEeqnarray}{rCl}
    \cI^{\tin} &=& \int \iota(\varepsilon(\kappa))  
    \lvert\varepsilon'(\kappa)\rvert \dif\kappa
    \label{eq:incidentenergyModel2}
    \\
    \cI^{\tout} &=& \int \iota(\varepsilon(\kappa))  
    \abs{\varepsilon'(\kappa)}  \E^{-\kappa \int \alpha(x,y) \dif\ell} 
    \dif\kappa
    \label{eq:polyModel2}
  \end{IEEEeqnarray}
\end{subequations}
Now, we define
\begin{equation}
  \label{eq:upiotaDef}
  \upiota(\kappa) \triangleq
  \iota(\varepsilon(\kappa)) |\varepsilon'(\kappa)|
\end{equation}
with $\varepsilon'(\kappa) = \dif \varepsilon(\kappa) / \dif \kappa$ and 
obtain \eqref{eq:massattenuationpar} by plugging \eqref{eq:upiotaDef} into 
\eqref{eq:massattenuationparametrization}.

As shown in Fig.~\ref{fig:triRelate}, the area 
$\iota(\varepsilon_j)\Delta\varepsilon_j$ depicting the X-ray energy within 
the $\Delta\varepsilon_j$ slot is the same as the corresponding area 
$\upiota(\kappa_j) \Delta\kappa_j$, the amount of X-ray energy that  
attenuated at a rate within $\Delta\kappa_j$ slot.  In the following, we 
extend the above results to the case where $\kappa(\varepsilon)$ is not 
monotonic.

\subsection{Piecewise-continuous $\kappa(\varepsilon)$ with monotonically 
decreasing invertible segments}
\label{app:piecewisecontinuouskappa}

\begin{figure}
  \centering
  \includegraphics{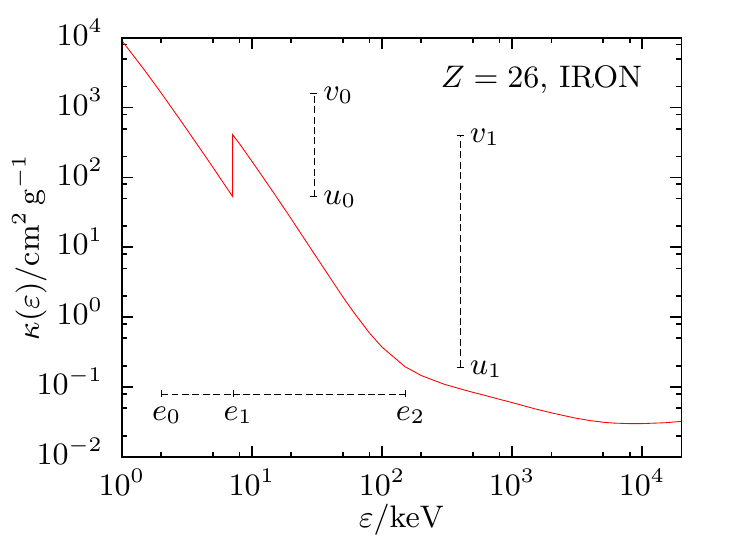}
  \caption{
    The mass attenuation coefficients $\kappa$ of iron versus the photon 
    energy $\varepsilon$ with a $K$-edge at 
    \SI{7.11}{\text{\kilo\electronvolt}}.
  }
  \label{fig:ironKappa}
\end{figure}

Define the domain $\mathbb{E}$ of $\varepsilon$ and a sequence of disjoint 
intervals $\{(e_m,e_{m+1})\}_{m=0}^M$ with $e_0=\min(\mathbb{E})$ and 
$e_{M+1}=\max(\mathbb{E})$, such that in each interval 
$\kappa(\varepsilon)$ is invertible and differentiable. Here, $\mathbb{E}$ 
is the support set of the incident X-ray spectrum $\iota(\varepsilon)$ and
$(e_m)_{m=1}^M$ are the $M$ $K$-edges in $\mathbb{E}$.  Taking 
Fig.~\ref{fig:ironKappa} as an example, there is only one $K$-edges at 
$e_1$ given the incident spectrum has its support as $(e_0,e_2)$.
  
The range and inverse of $\kappa(\varepsilon)$ within $(e_m, e_{m+1})$ are 
$(u_m, v_m)$ and $\varepsilon_m(\kappa)$, respectively, with
$u_m\df\inf_{\varepsilon{\scriptscriptstyle\nearrow}e_{m+1}}\kappa(\varepsilon)
<
v_m\df\sup_{\varepsilon{\scriptscriptstyle\searrow}e_m}\kappa(\varepsilon)$.  
Then, the noiseless measurement in \eqref{eq:polyModelseparable} can be 
written as
\begin{subequations}
  \begin{IEEEeqnarray}{rCl}
    \cI^{\text{out}} &=& \sum_{m=0}^M \int_{e_m}^{e_{m+1}}
    \iota(\varepsilon)
    \E^{- \kappa(\varepsilon) \int \alpha(x,y) \dif\ell} \dif\varepsilon\\
    &=& \sum_{m=0}^M \int_{u_m}^{v_m} \iota(\varepsilon_m(\kappa))
    \left| \varepsilon_m'(\kappa) \right|
    \E^{-\kappa
    \int \alpha(x,y) \dif\ell} \dif\kappa\\
    &=& \int \sum_{m=0}^M 1_{(u_m, v_m)}(\kappa)
    \iota(\varepsilon_m(\kappa)) \left| \varepsilon_m'(\kappa) \right| 
    \E^{-\kappa \int \alpha(x,y) \dif\ell}
    \dif\kappa. \label{eq:polyModel4}
  \end{IEEEeqnarray}
\end{subequations}
and  \eqref{eq:massattenuationpar} and \eqref{eq:generalUpiota} follow.  
Observe that \eqref{eq:generalUpiota} reduces to \eqref{eq:upiotaDef} when 
$M=0$.

In this case, suppose there is a $K$-edge with in the range 
$(\varepsilon_{j_1},\varepsilon_{j_2})$ and 
$\kappa(\varepsilon_{j_1})=\kappa(\varepsilon_{j_2})=\kappa_j$, then the 
amount of X-ray energy, $\upiota(\kappa_j) \Delta\kappa_j$, that attenuated 
at a rate $\kappa_j$ within $\Delta\kappa_j$ slot will be the summation of $\iota(\varepsilon_{j_1})\Delta\varepsilon_{j_1}$
and $\iota(\varepsilon_{j_2})\Delta\varepsilon_{j_2}$, where slot 
$\Delta\varepsilon_{j_1}$ and $\Delta\varepsilon_{j_2}$ correspond to 
$\Delta\kappa_j$.

\section{Proofs of Convexity of $\cL_\upiota(\balpha)$}
\label{app:biConvex}
\renewcommand{\theequation}{B\arabic{equation}}
\setcounter{equation}{0}

\newcommand*{\barmu}{\ensuremath{\bar{\mu}}}
\newcommand*{\bark}{\ensuremath{\bar{\kappa}}}
\newcommand*{\barw}{\ensuremath{\bar{w}}}

The proofs of convexity of $\cL_\upiota(\balpha)$ for both the lognormal 
and Poisson noise measurement models share a common component, which we 
first introduce as a lemma.

\begin{lem}
  \label{lemma:positiveBarw}
  For $\upiota(\kappa)$ that satisfy Assumption~\ref{assumption:I}, the 
  following holds:
  \begin{subequations}
    \label{eq:convLemma}
  \begin{IEEEeqnarray}{rCl}
    \label{eq:wdef}
    \tw &\df& \mathlarger{\iint}
    \biggl[\mu\kappa-\frac{q^{j_0}}{(q^{j_0}+1)^2}\PARENS{\mu+\kappa}^2\biggr]
    \upiota(\kappa)\upiota(\mu)h(\kappa+\mu)
    \dif\mu \dif\kappa \\
    &\geq& 0
  \end{IEEEeqnarray}
\end{subequations}
  for $q>1$ [as in \eqref{eq:q}] and any nonnegative function 
  $h:\mathbb{R}\rightarrow\mathbb{R}_+$.
\end{lem}

\begin{IEEEproof}
In Fig.~\ref{fig:muKappa}, the $(\mu,\kappa)$ coordinates of $P$, $B$ and 
$N$ are $(\kappa_0,0)$, $(\kappa_{j_0},0)$ and $(\kappa_{J+1},0)$, 
respectively;  the line $OS$ is defined by $\kappa=\mu$.

\begin{figure}
  \centering
  \includegraphics{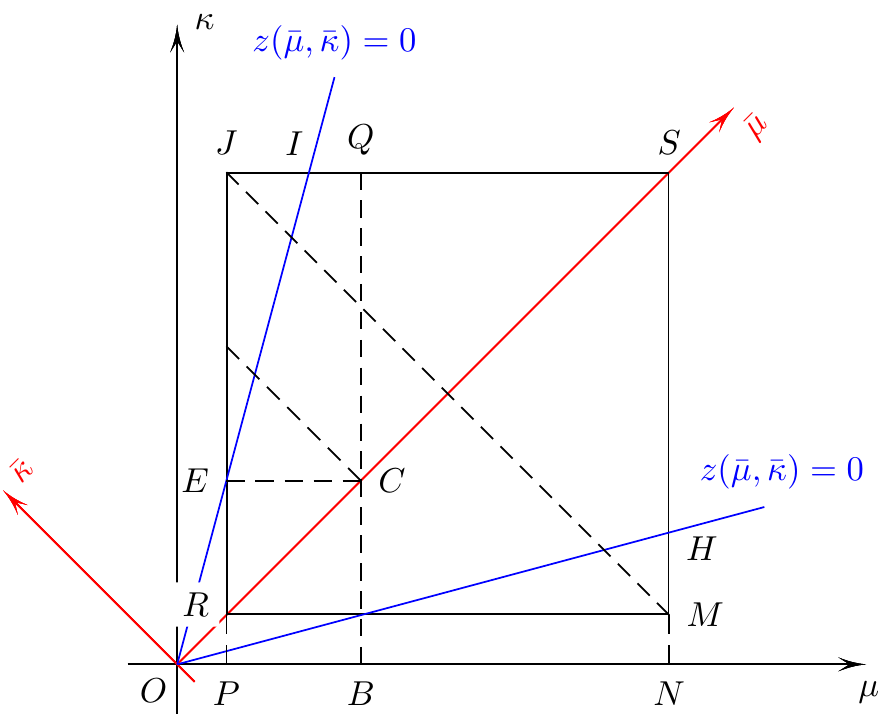}
  \caption{Integral region illustration.}
  \label{fig:muKappa}
\end{figure}

Considering the finite support set of $\iota(\kappa)$, the effective 
integral range is $[\kappa_0,\kappa_{J+1}]^2$, which is the rectangle 
$RMSJ$ in Fig.~\ref{fig:muKappa}.  Using the symmetry between $\kappa$ and 
$\mu$ in \eqref{eq:wdef}, we change the integral variables of 
\eqref{eq:wdef} by rotating the coordinates by $90^\circ$ using
\begin{subequations}
  \label{eq:rotateCoord}
  \begin{IEEEeqnarray}{rCl}
    \mu&=& \frac{1}{\sqrt{2}}(\barmu-\bark)
    \\
    \kappa&=& \frac{1}{\sqrt{2}}(\barmu+\bark)
  \end{IEEEeqnarray}
\end{subequations}
which yields
\begin{subequations}
  \begin{IEEEeqnarray}{rCl}
    \tw &=&
    \frac{1}{2}\int\displaylimits_{\sqrt{2}\kappa_0}^{\sqrt{2}\kappa_{J+1}}
    \int\displaylimits_{-g(\barmu)}^{g(\barmu)}
    \barw(\barmu,\bark)
    \dif\bark
      \: h\bigl(\sqrt{2}\barmu\bigr)
    \dif\barmu
    \\
    \label{eq:duetosymmetry}
    &=&
    \int\displaylimits_{\sqrt{2}\kappa_0}^{\sqrt{2}\kappa_{J+1}}
    \int\displaylimits_0^{g(\barmu)}
    \barw(\barmu,\bark)
      \dif\bark \:  h\bigl(\sqrt{2}\barmu\bigr)
\dif\barmu
  \end{IEEEeqnarray}
  where
  \begin{IEEEeqnarray}{rCl}
    \label{eq:barw}
    \barw(\barmu,\bark) &\df& z(\barmu,\bark)
    \upiota\PARENS{\frac{\barmu+\bark}{\sqrt{2}}}
    \upiota\PARENS{\frac{\barmu-\bark}{\sqrt{2}}}
    \\
    z(\barmu,\bark)&\df&  
    \PARENS{\frac{q^{j_0}-1}{q^{j_0}+1}}^2\barmu^2-\bark^2
    \\
    g(\barmu)&\df&
    \ccases{
      \barmu-\sqrt{2}\kappa_0,
      &
      \barmu\leq\frac{1}{\sqrt{2}}(\kappa_0+\kappa_{J+1})
      \\
      \sqrt{2}\kappa_{J+1}-\barmu,
      &
      \barmu>\frac{1}{\sqrt{2}}(\kappa_0+\kappa_{J+1})
    }
  \end{IEEEeqnarray}
\end{subequations}
and \eqref{eq:duetosymmetry} follows because \eqref{eq:barw} is even 
symmetric with respect to $\bark$.  Now, the integration region is reduced 
to the triangle $RSJ$ in Fig.~\ref{fig:muKappa}.

Note that $z(\barmu,\bark) \geq 0$ in the cone between lines $OH$ and $OI$, 
[both of which are specified by $z(\barmu,\bark)=0$], which implies that  
$\barw\PARENS{\barmu,\bark}\geq0$ within $RCE$ and $CSQ$; hence, the 
integrals of $\barw(\barmu,\bark) h\bigl(\sqrt{2}\barmu\bigr)$ over $RCE$ 
and $CSQ$ are nonnegative and, consequently,
\begin{IEEEeqnarray}{c}
  \label{eq:wineq}
  \tw \geq \iint\limits_\mathcal{R}^{} \barw(\barmu,\bark)
  \dif\bark \; h\bigl(\sqrt{2}\barmu\bigr) \dif\barmu.
\end{IEEEeqnarray}
where
\begin{IEEEeqnarray}{rCl}
  \mathcal{R} \df \CBR{
    (\barmu,\bark)\given{
      \frac{\barmu-\bark}{\sqrt{2}}\in[\kappa_0,\kappa_{j_0}],\;
      \frac{\barmu+\bark}{\sqrt{2}}\in[\kappa_{j_0},\kappa_{J+1}]
    }
  }
\end{IEEEeqnarray}
is our new integration region, which is the rectangle $ECQJ$ in 
Fig.~\ref{fig:muKappa}.

Define
\begin{IEEEeqnarray}{rCl}
  \label{eq:c}
  c \df \frac{\sqrt{2}}{1+q^{j_0}}.
\end{IEEEeqnarray}

Now, we split the inner integral over $\bark$ on the right-hand side of 
\eqref{eq:wineq} for fixed $\barmu$ into two regions: 
$z(\barmu,\bark)\geq0$ and $z(\barmu,\bark)<0$, i.e., trapezoid $ECQI$ and 
triangle $EIJ$, and prove that the positive contribution of the integral 
over $ECQI$ is larger than the negative contribution of the integral over 
the $EIJ$.

Note that the line $OI$ is specified by $z(\barmu,\bark)=0$ and the 
$(\mu,\kappa)$-coordinate of $I$ in Fig.~\ref{fig:muKappa} is thus 
$(\kappa_{J+1-j_0},\kappa_{J+1})$.
Hence, under Assumption~\ref{assumption:I}, region 
$ECQI\subseteq(\mathcal{K}_\text{low}\cup\mathcal{K}_\text{mid})\times\mathcal{K}_\text{high}$
and
region $EIJ\subseteq\mathcal{K}_\text{low}\times\mathcal{K}_\text{high}$.
Therefore, the following hold within $\mathcal{R}$:
\begin{itemize}
  \item When $z(\barmu,\bark) \geq 0$, i.e., in region $ECQI$,
    \begin{subequations}
      \label{eq:upiotaineq}
      \begin{IEEEeqnarray}{rCl}
        \label{eq:mpk}
        \lgiven{ \upiota(\kappa) }_{ \kappa=\frac{\barmu+\bark}{\sqrt{2}} } 
        &\geq&
        \upiota\PARENSbig{ c q^{j_0} \barmu} \\ \label{eq:mmk}
        \lgiven{ \upiota(\mu) }_{ \mu=\frac{\barmu-\bark}{\sqrt{2}} } &\geq&
        \upiota\PARENSbig{c \barmu}
      \end{IEEEeqnarray}
      where \eqref{eq:mpk} follows because 
      $\kappa=\frac{\barmu+\bark}{\sqrt{2}}$ takes values between 
      $\kappa_{j_0}$ and   $cq^{j_0}\barmu\in[\kappa_{j_0},\kappa_{J+1}]$, 
      i.e., $\kappa\in\mathcal{K}_\text{high}$ and $\upiota(\kappa)$ 
      decreases in $\mathcal{K}_\text{high}$;  \eqref{eq:mmk} follows 
      because $\mu=\frac{\barmu-\bark}{\sqrt{2}}$ takes values between 
      $c\barmu\in[\kappa_0,\kappa_{J+1-j_0}]$ and $\kappa_{j_0}$, i.e., 
      $\mu$ crosses $\mathcal{K}_\text{low}$ ($\upiota(\kappa)$ increasing) 
      and $\mathcal{K}_\text{mid}$ ($\upiota(\kappa)$ keeping high) 
      regions, see \eqref{eq:Kapparegions}.
      Here,  $(c\barmu, cq^{j_0}\barmu)$ is the $(\mu,\kappa)$-coordinate 
      of one point on line $OI$ specified by $\bar{\mu}$ in 
      $(\barmu,\bark)$-coordinate system.

    \item When $z(\barmu,\bark) < 0$, i.e., in region $EIJ$, 
      \begin{IEEEeqnarray}{rCl}
        \label{eq:mpk2}
        \lgiven{ \upiota(\kappa) }_{ \kappa=\frac{\barmu+\bark}{\sqrt{2}} } 
        &<&
        \upiota\PARENSbig{ c q^{j_0} \barmu} \\
        \label{eq:mmk2}
        \lgiven{ \upiota(\mu) }_{ \mu=\frac{\barmu-\bark}{\sqrt{2}} } &<&
        \upiota\PARENSbig{c \barmu}
      \end{IEEEeqnarray}
      where \eqref{eq:mpk2} follows because 
      $\kappa=\frac{\barmu+\bark}{\sqrt{2}}> c q^{j_0} \barmu$, i.e., 
      $\kappa \in \mathcal{K}_\text{high}$ and \eqref{eq:mmk2} follows 
      because $\mu=\frac{\barmu-\bark}{\sqrt{2}}< c \barmu$, i.e., $\mu \in 
      \mathcal{K}_\text{low}$.
    \end{subequations}
\end{itemize}

By combining \eqref{eq:upiotaineq} with \eqref{eq:wineq}, we have
\begin{subequations}
  \begin{IEEEeqnarray}{rCl}
    \label{eq:posNegComb}
    \tw &\geq&
    \iint_{\mathcal{R}}{
      z(\barmu,\bark)
      \dif\bark\;
      \bar{h}(\barmu)
    }
    \dif\barmu\\
    &=&  
    \int\limits_{\frac{\kappa_0+\kappa_{j_0}}{\sqrt{2}}}^{\frac{\kappa_{J+1}+\kappa_{J+1-j_0}}{\sqrt{2}}}
    \int\displaylimits_{\{\bark \,\mid\, (\barmu,\bark)\in\mathcal{R}\}}
    z(\barmu,\bark)
\dif \bark \dif \barmu
  \end{IEEEeqnarray}
  with
  \begin{IEEEeqnarray}{rCl}
    \label{eq:hbar}
    \bar{h}(\barmu) &\df&
\upiota\PARENSbig{ c q^{j_0} \barmu}
            \upiota\PARENSbig{c \barmu}
    h\bigl(\sqrt{2}\barmu\bigr) \notag\\
    &\geq& 0.
  \end{IEEEeqnarray}
\end{subequations}

It easy to verify that
$\int\displaylimits_{\{\bark \,\mid\, (\barmu,\bark)\in\mathcal{R}\}}
  z(\barmu,\bark) \dif\bark$
is an increasing function of $\barmu$ over the range of the outer integral 
$\SBR{\frac{\kappa_0+\kappa_{j_0}}{\sqrt{2}},\frac{\kappa_{J+1}+\kappa_{J+1-j_0}}{\sqrt{2}}}$ 
and, consequently,
\begin{IEEEeqnarray}{c}
  \label{eq:ineqlemma}
  \int\displaylimits_{\{\bark \,\mid\, (\barmu,\bark)\in\mathcal{R}\}}
  z(\barmu,\bark) \dif\bark\geq0
\end{IEEEeqnarray}
where the equality is attained for 
$\barmu=(\kappa_0+\kappa_{j_0})/\sqrt{2}$.  Finally, \eqref{eq:convLemma} 
follows from \eqref{eq:posNegComb} and \eqref{eq:ineqlemma}.
\end{IEEEproof}

This proof of convexity of Lemma~\ref{lemma:positiveBarw} is 
conservative as we loose the positve integral in region $RCE$ and $CSQ$ to 
zero.

Now, we use Lemma~\ref{lemma:positiveBarw} to prove the convexity of 
$\cL_\upiota(\balpha)$ in Lemmas~\ref{lemma:alphaconvexlognormal} and 
\ref{lemma:alphaconvexPoisson}.  Note that the mass-attenuation spectrum 
$\upiota(\kappa)$ is considered known in 
Lemmas~\ref{lemma:alphaconvexlognormal} and \ref{lemma:alphaconvexPoisson} 
and define $\xi(\cdot)\df\upiota^\tL(\cdot)$ and the corresponding first 
and second derivatives:
\begin{IEEEeqnarray}{rCl"rCl}
  \label{eq:xiDef}
  \dot{\xi}(s) &=& \PARENS{-\kappa\upiota}^\tL(s), &
  \ddot{\xi}(s) &=& \PARENS{\kappa^2\upiota}^\tL(s).
\end{IEEEeqnarray}
Observe that $\bcI^\tout = \PARENS{\cI^\tout_n}_{n=1}^N =  
\xi_\circ(\Phi\balpha) = \PARENSbig{\xi(\bphi^T_n\balpha)}_{n=1}^N$.  For 
notational simplicity, we omit the dependence of $\bcI^\tout$ on $\balpha$ 
and $\bcI$ and use $\bcI^\tout$ and $\xi_\circ(\Phi\balpha)$ 
interchangeably. 

\begin{IEEEproof}[Proof of Lemma~\ref{lemma:alphaconvexlognormal}]
We use the identities
 \begin{subequations}
  \label{eq:dersofi}
  \begin{IEEEeqnarray}{rCl}
    \frac{\partial\xi_\circ(\Phi\balpha)}{\partial\balpha^T} &=& 
    \diag\PARENSbig{ \dot{\xi}_\circ(\Phi\balpha)  } \Phi\\
    \frac{\partial\xi(\bphi^T_n\balpha)}{\partial\balpha\partial\balpha^T} 
    &=&
    \ddot{\xi}(\bphi^T_n\balpha) \bphi_n\bphi_n^T
  \end{IEEEeqnarray}
\end{subequations}
to compute the gradient and Hessian
of the lognormal \gls{NLL} in \eqref{eq:lognormalNLLalpha}:
\begin{subequations}
  \begin{IEEEeqnarray}{rCl}
    \frac{\partial \cL_\upiota(\balpha)}{\partial\balpha}
    &=& \Phi^T\diag\PARENSbig{\dot{\xi}_\circ(\Phi\balpha)}
    \diag^{-1}\PARENS{\bcI^\tout}
    \PARENS{\ln_\circ\bcI^\tout-\ln_\circ\bcE}\\
    \label{eq:lognormalHessianA}
    \frac{\partial \cL_\upiota(\balpha)}{\partial\balpha\partial\balpha^T}
    &=& \Phi^T \diag^{-2}\PARENS{\bcI^\tout}\diag\PARENS{\bw}\Phi
  \end{IEEEeqnarray}
  where the $N \times 1$ vector $\bw = \PARENS{w_n}_{n=1}^N$ is defined as
  \begin{IEEEeqnarray}{rCl}
    \label{eq:wn}
    w_n &=& \lgiven{ \dot{\xi}^2(s) + t(s) \PARENS{\ln \cI_n^\tout - \ln 
    \cE_n} }_{s=\bphi_n^T\balpha}  \\
    t(s) &\df&   \xi(s) \ddot{\xi}(s) - \dot{\xi}^2(s).
  \end{IEEEeqnarray}
\end{subequations}

For $s>0$,
\begin{subequations}
\begin{IEEEeqnarray}{rCl}
  \label{eq:cauchyschwarz}
  t(s) &=& \int \upiota(\kappa) \E^{-s \kappa} \dif \kappa \int \kappa^2 
  \upiota(\kappa) \E^{-s \kappa} \dif \kappa - \biggl[\int \kappa 
  \upiota(\kappa) \E^{-s \kappa} \dif \kappa\biggr]^2 \geq 0
\end{IEEEeqnarray}
where we used the Laplace-transform identity for derivatives 
\eqref{eq:LTders} and
the inequality follows by using the Cauchy-Schwartz inequality.  We can 
write $t(s)$ as 
\begin{IEEEeqnarray}{rCl}
  \label{eq:tsSim}
  t(s)  =  \frac{1}{2} \iint 
  (\kappa-\mu)^2\upiota(\kappa)\upiota(\mu)\E^{-s(\kappa+\mu)} \dif \kappa 
  \dif \mu
\end{IEEEeqnarray}
\end{subequations}
which also implies that $t(s) \geq 0$.  We obtain \eqref{eq:tsSim}
by using the Laplace-transform identity for derivatives \eqref{eq:LTders}, 
combining the multiplication of the integrals,
and symmetrizing the integral expressions by replacing $\kappa^2$ with 
$(\kappa^2+\mu^2)/2$.

According to \eqref{eq:Lalphaconvexcond}, $\ln\cI^\tout_n-\ln\cE_n\geq-U$, 
see also \eqref{eq:U}.  Because of positivity of $t(s)$ for $s\geq0$,
\begin{subequations}
\begin{IEEEeqnarray}{rCl}
  \label{eq:wn1}
  w_n &\geq&  \lgiven{ \dot{\xi}^2(s) -U t(s)
  }_{s=\bphi_n^T\balpha} \\
  \label{eq:wn2}
  &=& \mathlarger{\iint}
  \biggl[\mu\kappa - U\frac{(\mu-\kappa)^2}{2} \biggr]
  \upiota(\mu)\upiota(\kappa)
  \E^{-(\mu+\kappa)\bphi_n^T\balpha}
  \dif\kappa \dif\mu\\
  &=&
  \label{eq:plugInU}
  \PARENS{\frac{q^{j_0}+1}{q^{j_0}-1}}^2
  \mathlarger{\iint}
  \biggl[\mu\kappa - \frac{q^{j_0}}{(q^{j_0}+1)^2}(\mu+\kappa)^2 \biggr]
  \upiota(\mu)\upiota(\kappa)
  \E^{-(\mu+\kappa)\bphi_n^T\balpha}
  \dif\kappa \dif\mu
\end{IEEEeqnarray}
\end{subequations}
where \eqref{eq:wn2} is obtained by plugging \eqref{eq:tsSim} into 
\eqref{eq:wn1}, and we have used the definition of $U$ \eqref{eq:U} in  
\eqref{eq:plugInU}.

Using Lemma~\ref{lemma:positiveBarw} with 
$h(\kappa)=\E^{-\kappa\bphi_n^T\balpha}$, we have $w_n\geq0$ for all $n$.  
Therefore, the Hessian matrix \eqref{eq:lognormalHessianA} of 
$\cL_\upiota(\balpha)$ is positive semidefinite.
\end{IEEEproof}

\begin{IEEEproof}[Proof of Lemma~\ref{lemma:alphaconvexPoisson}] We use the 
  identities \eqref{eq:dersofi} to compute the gradient and Hessian of the  
  Poisson \gls{NLL} \eqref{eq:poissonNLLalpha}:
\begin{subequations}
  \begin{IEEEeqnarray}{rCl}
    \frac{\partial \cL_\upiota(\balpha)}{\partial\balpha}
    &=& \Phi^T\diag\PARENSbig{\dot{\xi}_\circ\PARENSbig{\Phi\balpha}}
    \SBRbig{\bm{1}-\diag^{-1}\PARENS{\bcI^\tout}\bcE}
    \\
    \label{eq:poissHeissA}
    \frac{\partial \cL_\upiota(\balpha)}{\partial\balpha\partial\balpha^T}
    &=& \Phi^T 
    \diag^{-2}\PARENS{\bcI^\tout}\diag\PARENS{\bcE}\diag\PARENS{\bx}\Phi
  \end{IEEEeqnarray}
where the $N \times 1$ vector $\bx = \PARENS{x_n}_{n=1}^N$ is defined as
  \begin{IEEEeqnarray}{rCl}
  \label{eq:poissWi}
  x_n &=& \lgiven{ \dot{\xi}^2\PARENSbig{s}
  +\ddot{\xi}\PARENSbig{s}\xi(s)
  \PARENS{\frac{\cI^\tout_n}{\cE_n}-1}}_{s=\bphi_n^T\balpha}.
  \end{IEEEeqnarray}
\end{subequations}
Since $\cI_n^\tout\geq (1-V)\cE_n\geq 0$ according to 
\eqref{eq:poissonConvCondalpha}, we have
\begin{equation}
  \frac{\cI^\tout_n}{\cE_n}-1\geq-V.
  \label{eq:VjInequality}
\end{equation}

Now,
\begin{subequations}
  \begin{IEEEeqnarray}{rCl}
    \label{eq:xn}
    x_n &\geq& \iint \PARENS{ \mu\kappa - \kappa^2V }
    \upiota(\mu)\upiota(\kappa)
    \E^{-(\mu+\kappa)\bphi_n^T\balpha}
    \dif\kappa\dif\mu\\
    \label{eq:poissSymmetry}
    &=& {\iint}
    \Bigl({\mu\kappa - \frac{\mu^2+\kappa^2}{2}V}\Bigr)
    \upiota(\mu)\upiota(\kappa)
    \E^{-(\mu+\kappa)\bphi_n^T\balpha}
    \dif\kappa\dif\mu
    \\
  &\geq&
    \frac{(q^{j_0}+1)^2}{q^{2j_0}+1}
    \mathlarger{\iint}
    \Bigl[\mu\kappa-\frac{q^{j_0}}{(q^{j_0}+1)^2}(\mu+\kappa)^2\Bigr]
    \upiota(\mu)\upiota(\kappa)
    \E^{-(\mu+\kappa)\bphi_n^T\balpha}
    \dif\kappa\dif\mu
    \\
    \label{eq:xnonneg}
    &\geq& 0
  \end{IEEEeqnarray}
\end{subequations}
where \eqref{eq:xn} follows by applying inequality \eqref{eq:VjInequality} 
to \eqref{eq:poissWi}, using the Laplace-transform identity for derivatives 
\eqref{eq:LTders}, and combining the multiplication of the integrals,  and 
\eqref{eq:poissSymmetry} is due to the symmetry with respect to $\mu$ and 
$\kappa$ by replacing $\kappa^2$ with $(\kappa^2+\mu^2)/2$.  Now, plug 
\eqref{eq:V} into \eqref{eq:poissSymmetry} and apply 
Lemma~\ref{lemma:positiveBarw} with 
$h(\kappa)=\E^{-\kappa\bphi_n^T\balpha}$ to conclude \eqref{eq:xnonneg}.  
Therefore, the Hessian of $\cL_\upiota(\balpha)$ in   
\eqref{eq:poissHeissA} is positive semidefinite.
\end{IEEEproof}

\section{Proofs of \gls{KUL} Property and Convergence of PG-BFGS Iteration}
\label{app:KL_PropertyProofConv}

\renewcommand{\theequation}{C\arabic{equation}}
\setcounter{equation}{0}

\begin{IEEEproof}[Proof of Theorem~\ref{th:kl}]
According to \cite{XuYin2013}, real-analytic function \cite{Rudin1976}, 
semialgebraic functions \cite{BochnakCosteRoy} and their summations satisfy 
the \gls{KUL} property automatically.  Therefore, the proof consists two 
parts:
\begin{enumerate}[label=(\alph*)]
  \item\label{realAnalytic}
    The \glspl{NLL} in \eqref{eq:lognormalNLL} and \eqref{eq:poissNLL}, are 
    both real-analytic functions of $(\balpha,\bcI)$ on 
    $\mathbb{C}\subseteq\dom(f)$;
  \item\label{semialgebraic}
    Both $r(\balpha)$ in \eqref{eq:rWV}--\eqref{eq:rTV} and 
    $\mathbb{I}_\nonneg(\bcI)$ are semialgebraic functions;
\end{enumerate}
and, consequently, the objective function $f(\balpha,\bcI)$ satisfies the 
\gls{KUL} property.

\textbf{Real-analytic \glspl{NLL}.}
The \glspl{NLL} in \eqref{eq:lognormalNLL} and \eqref{eq:poissNLL} are in 
the form of weighted summations of terms $\bb^\tL(\bphi_n^T\balpha)\bcI$, 
$\ln\SBRbig{\bb^\tL(\bphi_n^T\balpha)\bcI}$, and 
$\ln^2\SBRbig{\bb^\tL(\bphi_n^T\balpha)\bcI}$ for $n=1,2,\ldots,N$.  
Weighted summation of real-analytic functions is real-analytic; hence, we 
need to prove that the following are real-analytic functions:
\begin{subequations}
  \label{eq:buildingblock}
  \begin{IEEEeqnarray}{rCl}
    l_1(t) &=&
    \bb^\tL\PARENSbig{\bphi^T(\balpha+t\bgamma)}(\bcI+t\bcJ)
    \\
    l_2(t) &=& \ln\SBRbig{ 
      \bb^\tL\PARENSbig{\bphi^T(\balpha+t\bgamma)}(\bcI+t\bcJ)
    }=\ln l_1(t) \\
    l_3(t) &=& l_2^2(t)
  \end{IEEEeqnarray}
\end{subequations}
Since $\PARENS{l_i(t)}_{i=1}^3$ are smooth, it is sufficient to prove that 
the $m$th derivatives, $l_i^{(m)}(t)$, are bounded for all $m$, $(\balpha, 
\bcI)$, $(\bgamma, \bcJ)$ and $t$ such that $(\balpha+t\bgamma, 
\bcI+t\bcJ)\in\dom(f)$.
The $m$th derivative of  $l_1(t)$ is
\begin{equation}
  l_1^{(m)}=(\bphi^T\bgamma)^m
 \PARENS{(-\kappa)^m\bb}^\tL (\balpha+t\bgamma)
  (\bcI+t\bcJ)
  +m(\bphi^T\bgamma)^{m-1}
\PARENS{(-\kappa)^{m-1}\bb}^\tL (\balpha+t\bgamma) \bcJ
  \label{eq:l1m}
\end{equation}
which is bounded for any $\balpha$, $\bcI$, $\bgamma$, $\bcJ$ and $t$ such 
that $(\balpha+t\bgamma, \bcI+t\bcJ)$ is in one of compact subsets 
$\mathbb{C}\subseteq\dom(f)$.
For any compact set $\mathbb{C}\subseteq\dom(f)$, there exists $\epsilon>0$ 
such that $l_1(t)\geq\epsilon$, for all $(\balpha+t\bgamma, 
\bcI+t\bcJ)\in\mathbb{C}$. Hence, $\ln(\cdot)$ and $(\cdot)^2$ are analytic 
on $[\epsilon,+\infty)$.  Now, since the compositions and products of 
  analytic functions are analytic \cite[Ch.~1.4]{Krantz2002}, we have that 
  both $l_2(t)$ and $l_3(t)$ are analytic.

Therefore, the \glspl{NLL} in both \eqref{eq:lognormalNLL} and 
\eqref{eq:poissNLL} are analytic.

\textbf{Semialgebraic regularization terms.}
According to \cite{XuYin2013},
\begin{enumerate*}[label=\roman*)]
  \item the $\ell_2$ and $\ell_1$ norms $\norm{\cdot}_2$ and 
    $\norm{\cdot}_1$ are semialgebraic,
  \item indicator function $\mathbb{I}_\nonneg(\cdot)$ is semialgebraic,
  \item finite sums and products of semialgebraic functions are 
    semialgebraic, and
  \item the composition of semialgebraic functions are semialgebraic.
\end{enumerate*}
Therefore, $\norm{\Psi^T\balpha}_1$, $\mathbb{I}_\nonneg(\balpha)$ and 
$\mathbb{I}_\nonneg(\bcI)$ are all semialgebraic.  Since 
$\sqrt{\sum_{j\in\mathcal{N}_i}(\alpha_i-\alpha_j)^2}=\normlr{P_i\balpha}_2$ 
for some matrix $P_i$ is semialgebraic, $r(\balpha)$ in \eqref{eq:rTV} is 
semialgebraic.

Finally, according to \cite{XuYin2013}, sum of real-analytic and 
semialgebraic functions satisfies \gls{KUL} property. Therefore, 
$f(\balpha,\bcI)$ satisfies the \gls{KUL} property on $\dom(f)$.
\end{IEEEproof}

\begin{IEEEproof}[Proof of Theorem~\ref{th:conv}]
We apply \cite[Lemma~2.6]{XuYin2013} to establish the convergence of 
$\CBRbig{\PARENSbig{\step{\balpha}{i},\step{\bcI}{i}}}_{i=1}^{+\infty}$. 

Since \eqref{eq:lognormalNLL}, $r(\balpha)$ in \eqref{eq:r} and 
$\mathbb{I}_\nonneg(\bcI)$ are lower-bounded, we only need to prove that 
$\eqref{eq:poissNLL}$ is lower-bounded.  By using the fact that $\ln x\leq 
x-1$, we have
\begin{subequations}
  \label{eq:NLLPoisGeq0}
  \begin{IEEEeqnarray}{rCl}
    \cL(\balpha,\bcI)
    &=&\bm{1}^T\SBR{\bcI^\tout(\balpha,\bcI)-\bcE}
    -\sum_{n=1}^N\cE_n\ln\frac{\cI_n^\tout(\balpha,\bcI)}{\cE_n}\\
    &\geq& 0
  \end{IEEEeqnarray}
\end{subequations}

According to the assumption, we have $f(\balpha,\bcI)$ strongly convex over 
$\bcI$ and bounded the step size $\beta^{(i)}$.
So, there exist constants $0<l<L< +\infty$ such that
\begin{subequations}
  \begin{IEEEeqnarray}{rCl}
    f\PARENSbig{\balpha^{(i+1)},\bcI^{(i)}}-f\PARENSbig{\balpha^{(i+1)},\bcI^{(i+1)}} 
    &\geq&
    \frac{l}{2}\normbig{\bcI^{(i)}-\bcI^{(i+1)}}_2^2\\
    L &\geq& \frac{1}{\beta^{(i)}} \geq l.
  \end{IEEEeqnarray}
\end{subequations}
Hence, we have verified all conditions of \cite[Lemma~2.6]{XuYin2013}.  
\end{IEEEproof}

\section*{Acknowledgment}

The authors are grateful to Dr.\ Joseph~N.\ Gray, Center for
Nondestructive Evaluation, Iowa State University, for providing real
X-ray CT data used in the numerical examples in 
Section~\ref{sec:realXrayCTex}. 

\printbibliography
\end{document}